\documentclass[a4paper, 11pt]{article}
\usepackage{}

\usepackage{amsfonts}
\usepackage{dsfont}
\usepackage{mathrsfs}
\usepackage{amssymb}
\usepackage{bbm}
\usepackage{epsfig}
\usepackage{amsmath}
\usepackage{appendix}
\usepackage{float}
\usepackage{color}
\usepackage[english]{babel}

\newtheorem{thm}{Theorem}
\newtheorem{defn}{Definition}

\newtheorem{lem}{Lemma}
\newtheorem{remark}{Remark}
\newtheorem{prop}{Proposition}

\title{\bf   Multi-agent Robust Consensus\\ Convergence Analysis and Application\footnote{This work has been supported in part
by the Knut and Alice Wallenberg Foundation and the Swedish Research
Council.}}
\date{}

\author{Guodong Shi and Karl Henrik Johansson\thanks{G. Shi and K. H. Johansson are with ACCESS Linnaeus Centre, School of Electrical Engineering,
Royal Institute of Technology, Stockholm 10044, Sweden.
       Email: {\tt\small guodongs@kth.se, kallej@ee.kth.se}}
}

\begin{document}

\maketitle

\begin{abstract}
The paper investigates consensus problem for continuous-time multi-agent systems with time-varying communication graphs subject to process noises. Borrowing the ideas from input-to-state stability (ISS) and integral input-to-state stability (iISS), robust consensus and integral robust consensus are defined with respect to $L_\infty$ and $L_1$ norms of the disturbance functions, respectively. Sufficient and/or necessary connectivity conditions are obtained for the system to reach robust consensus or integral robust consensus, which answer the question: how much communication capacity is required for a multi-agent network to converge  despite certain amount of disturbance. The $\epsilon$-convergence time is then obtained for the network as a special case of the robustness analysis. The results are based on quite general assumptions on switching graph, weights rule and noise regularity.  In addition, as an illustration of the applicability of the results, distributed event-triggered coordination is studied.
\end{abstract}


\section{Introduction}
Coordination of multi-agent networks has attracted a significant amount of attention in the past few years, due to its broad applications in various fields of science including physics, engineering, biology, ecology and social science \cite{vic95, jad03, saber04,
mor,fax}. Distributed control using neighboring information flow has been shown to ensure collective tasks  such as formation, flocking, rendezvous, and aggregation \cite{sabertac, lin07, mar, shi09}.

Central to  multi-agent coordination study is the study of consensus, or state agreement, which requires that all agents achieve the desired relative position and the same velocity. Consensus seeking is extensively studied in the literature for both continuous-time and discrete-time models  \cite{tsi,  jad03, mor, tantac, tan04,caoming1, caoming2, ren05, lin07, lwang,xie}. Recently also asynchronous event-triggered sampling for such consensus-seeking multi-agent system was studied  \cite{lemmon, dimos, george}.  Researchers are not only concerned with what connectivity conditions can guarantee consensus, but also with the convergence rate: how fast the network reaches a consensus under certain connectivity assumptions \cite{lin07,caoming2,tsi1,tsi2, boyd}.

{  Consensus algorithms are usually carried out over an underlying communication network. Thus, connectivity of this  communication graph plays a key role in consensus analysis.} Various connectivity conditions have been used to describe frequently switching topologies in different cases. The ``joint connection", i.e., the union of graphs over a time interval, and
similar concepts are important in the analysis of consensus stability with
time-dependent topology. Uniformly joint connectedness, which requests the joint connection is connected for all intervals which are longer than some positive constant,  has been employed for different consensus problems from discrete-time to continuous-time agent dynamics, from directed to undirected interconnection topologies
\cite{tsi, jad03, lin07, hong07, cheng}. \cite{tsi} studied the
distributed asynchronous iterations, while \cite{jad03} proved the
consensus of a simplified Vicsek model.  Furthermore, \cite{hong07}
and \cite{cheng} investigated the jointly-connected coordination for
second-order agent dynamics.  A nonlinear continuous-time model was discussed in \cite{lin07}  with
directed communications, in which convergence to a consensus is shown to be uniform within bounded initial conditions. On the other hand,    $[t, \infty)$-joint connection requires the joint connection is connected for infinitely many disjoint intervals in $[0,\infty]$, which was discussed in \cite{mor} for consensus seeking of discrete-time agents. This connectivity concept was then
extended in continuous-time distributed control analysis for target set
convergence and state agreement in \cite{shi09}.

Communication over networks is typically unreliable and with channel noise, which has attracted researchers to look at the robustness of consensus algorithms \cite{wanglin1, wanglin2, munz, qhui, Tian}. In \cite{pesco}, robustness performance was discussed for average consensus algorithms. Then in \cite{wanglin1,wanglin2}, robust consensus was studied under directed communication graphs for discrete-time systems. For continuous-time multi-agent systems, robustness of consensus  was established by an $H_2$ bound for networks of single integrators with a fixed directed communication graph \cite{leonard}. In \cite{Tian}, robust consensus  with diverse input delays and
asymmetric interconnection perturbations was discussed for a second-order leader-following model.  Recently, an optimal synchronization protocol was studied for discrete-time double integrators subject to process noise \cite{carli}.

 Clearly, robustness of consensus algorithms subject to noise highly relies on the convergence rate for the algorithm in the absence of noise. The concept of $\epsilon$-convergence  time, was introduced to quantify the convergence speed for discrete-time consensus algorithms. It is defined as the minimum time steps required  for the network to reach a certain level of consensus captured by a parameter $\epsilon$.  Bounds of $\epsilon$-convergence  time have been widely established  in the literature for first-order discrete-time dynamics \cite{boyd, caoming2, tsi1, tsi2}, and  recently a sharp bound was presented in \cite{tsi2} indicating that the convergence time  is of order $O(n^2B)$, where $n$ is the number of nodes in the network and $B$ is a lower bound for the  time interval in the definition of uniformly joint connectedness. Few results have been obtained on the convergence rates for continuous-time multi-agent systems reaching a consensus with general joint connectivity assumptions. The robustness consensus analysis for continuous-time systems a challenging problem. A quantitative answer to  how much noise can be dealt with by how much communication is still missing.

The primary aim of this paper is to establish the convergence  towards a consensus for first-order, continuous-time multi-agent systems with communication noise for general directed and time-varying interconnection graphs.  Borrowing ideas from input-to-state stability (ISS) and integral input-to-state stability (iISS) \cite{sontag1, sontag-wang}, we define robust consensus and integral robust consensus for  the relative-position-based continuous-time coordination protocol which was first introduced  in \cite{saber04}.  We present explicit convergence bounds for the system with respect to $L_\infty$ and $L_1$ norms of the disturbances. Sufficient and necessary connectivity conditions are obtained for the system to reach robust consensus or integral robust consensus, respectively, for directed and bidirectional communications.  Consequently, the upper bounds for the $\epsilon$-convergence  time are established under uniform or non-uniform joint connections, as a simple special case of the robust consensus analysis.   To the best of our knowledge, the results from this analysis are the first to show that consensus is reached exponentially in $t$ with uniformly joint connected graphs, while ``exponentially" in the times that the joint graph are connected with $[t, \infty)$-joint connection for the considered systems. Different with most existing work, we build the whole analysis on a generalized integral assumption for the weight functions, which covers a large amount of common functions.  Additionally, a class of event-triggered coordination rules is studied as the application of the robust consensus results.

The paper is organized as follows. In Section 2, some preliminary concepts are introduced. We set up the system model, present our standing assumptions and main results in Section 3. Then convergence analysis is carried out for directed and bidirectional graphs in Sections 4 and 5, respectively.   We turn to event-triggered consensus in Section 6, as an application of the results obtained in previous sections. Finally, concluding remarks are given in Section 7.

%%%%%%%%%%%%%%%%%%%%%%%%%%%%%%%%%%%%%%%%%%%%%%%%%%%%%%%%%%%%%%%%%%%%%%%%%%%%%%%%%%%%%%%%%%%%%%%%%%%%%%%%%%%%%%%%%%%%%%%%%%%%%%%%%%%%%%%%%%%%
\section{Preliminaries}
Here we introduce some notations and theories on directed graphs and Dini derivatives.

\subsection{Directed Graphs} A directed graph (digraph) $\mathcal
{G}=(\mathcal {V}, \mathcal {E})$ consists of a finite set
$\mathcal{V}$ of nodes and an arc set
$\mathcal {E}$ \cite{god}.  An element $e=(i,j)\in\mathcal {E}$ is called an
{\it arc}  from node $i\in \mathcal{V}$  and to node $j\in\mathcal{V}$. If the
arcs are pairwise distinct in an alternating sequence
$ v_{0}e_{1}v_1e_{2}v_{2}\dots e_{n}v_{n}$ of nodes $v_{i}$ and
arcs $e_{i}=(v_{i-1},v_{i})\in\mathcal {E}$ for $i=1,2,\dots,n$,
the sequence  is called a (directed) {\it  path} with {\it length} $n$, and if $v_0=v_n$ a
(directed) {\it  cycle}. A path with no repeated nodes is called a {\it simple path}. A digraph without cycles is
said to be {\it acyclic}.

A path from $i$ to
$j$ is denoted as $i \rightarrow j$, and the length of $i \rightarrow j$ is denoted as $|i \rightarrow j|$.  If there exists a path from node $i$ to node $j$,
then node $j$ is said to be reachable from node $i$.  Each node is thought to be reachable by itself. A node $v$ from which any other node is
reachable  is called a {\it center} (or a {\it root}) of $\mathcal {G}$.  A digraph $\mathcal
{G}$ is said to be {\it strongly connected}  if it contains path $i \rightarrow j$ and $j \rightarrow i$ for every pair of nodes $i$ and $j$;   {\it
quasi-strongly connected} if $\mathcal {G}$ has a center
\cite{ber, lin07}.

In this paper, we define the (generalized) {\it distance} from $i$ to $j$, $d(i,j)$,   as the length of a longest simple path $i \rightarrow j$ if $j$ is reachable from $i$, and the (generalized) {\it diameter} of $\mathcal
{G}$ as $\max\{d(i,j)|i,j \in\mathcal
{V},\ j\mbox{ is reachable from}\ i\}$.

A digraph $\mathcal {G}$ is said to be bidirectional if for every two nodes $i$ and $j$, $i$ is a neighbor of $j$ if and only if $j$ is a neighbor of
$i$. Then {\it path, distance, diameter} can be similarly defined for $\mathcal{G}$ by ignoring the direction of the arcs. A bidirectional graph $\mathcal{G}$ is said to be {\it connected} if there is a path between any two nodes.

\subsection{Dini Derivatives}  The upper Dini
derivative of  a function $h: (a,b)\to \mathds{R}$ at $t$ is defined as
$$
D^+h(t)=\limsup_{s\to 0^+} \frac{h(t+s)-h(t)}{s}
$$
 The next
result is useful for the calculation of Dini derivatives
 \cite{dan,lin07}.

\begin{lem}
\label{lem1}  Let $V_i(t,x): \mathds{R}\times \mathds{R}^m \to \mathds{R}\;(i=1,\dots,n)$ be
$C^1$ and $V(t,x)=\max_{i=1,\dots,n}V_i(t,x)$. If $
\mathcal{I}(t)=\{i\in \{1,2,\dots,n\}\,:\,V(t,x(t))=V_i(t,x(t))\}$
is the set of indices where the maximum is reached at $t$, then
$
D^+V(t,x(t))=\max_{i\in\mathcal{ I}(t)}\dot{V}_i(t,x(t)).
$
\end{lem}

\vspace{2mm}

{\it Notations:} For a vector $z=(z_1,\dots,z_N)^T$ in $\mathds{R}^N$, $|z|$ denotes the maximum norm, i.e., $|z|\doteq \max_{i=1,\dots,N}|z_i|$. When $z: \mathds{R}_{\geq0}\rightarrow \mathds{R}^N$ is a measurable function defined on $[0,+\infty)$, $\|z\|_\infty$ denotes the essential supremum of $\{|z(t)|,t\in[0,+\infty)\}$. Moreover, a function $\gamma: R_{\geq 0}\rightarrow R_{\geq 0}$ is said to be
a $\mathcal{K}$-class function if it is continuous, strictly
increasing, and $\gamma(0)=0$;  a function $ \beta:\mathds{R}_{\geq
0}\times \mathds{R}_{\geq 0}\rightarrow \mathds{R} $ is a $\mathcal {KL}$-class function
if $\beta(\cdot,t)$ is of class $\mathcal {K}$ for each fixed $t\geq
0$ and $\beta(s,t)\rightarrow 0$ decreasingly as $t\rightarrow\infty$ for
each fixed $s\geq 0$.

%%%%%%%%%%%%%%%%%%%%%%%%%%%%%%%%%%%%%%%%%%%%%%%%%%%%%%%%%%%%%%%%%%%%%%%%%%%%%%%%%%%%%%%%%%%%%%%%%%%%%%%%%%%%%%%%%%%%%%%%%%%%%%%%%%%%%%%%%%%%
\section{Problem Statement and Main Results}
This paper considers a multi-agent system with agent set $\mathcal
{V}=\{1,\dots,N\}$, $N\geq2$, for which the dynamics of each agent is a
 first-order integrator:
\begin{equation}\label{0}
\dot{x}_i=u_i, \quad i=1,\dots,N
\end{equation}
where $x_i\in \mathds{R}$ represents the state of agent $i$, and $u_i$ is
its control input. Let $x=(x_1,\dots,x_N)^T$.

\subsection{Network, Dynamics and Assumptions}
The communication in the network is modeled as a time-varying graph $\mathcal {G}_{\sigma(t)}=(\mathcal {V},\mathcal
{E}_{\sigma(t)})$ with
$\sigma:[0,+\infty)\rightarrow \mathcal {Q}$  a piecewise constant function,
and  $\mathcal {Q}$ a finite set indicating all possible graphs. Node $j$ is said to be a {\it neighbor} of $i$ at time $t$ when there is an arc $(j,i)\in \mathcal
{E}_{\sigma(t)}$. Let $\mathcal {N}_i(\sigma(t))$ represent the set of agent $i$'s neighbors at time $t$. An assumption is given to the variation of $\mathcal {G}_{\sigma(t)}$.

\noindent {\bf A1.} {\it (Dwell Time)} There is a lower bound constant $\tau_D>0$ between two consecutive
switching time instants of $\sigma(t)$.

Denote the joint graph of $\mathcal
{G}_{\sigma(t)}$ in
time interval $[t_1,t_2)$ with $t_1<t_2\leq +\infty$ as
$\mathcal {G}\big([t_1,t_2)\big)= \big(\mathcal {V},\cup_{t\in[t_1,t_2)}\mathcal
{E}_{\sigma(t)}\big)$. We introduce the following definitions on connectivity.

\begin{defn}(i) $\mathcal
{G}_{\sigma(t)}$ is said to be {\it uniformly (jointly) strongly connected} (USC) if there exists a constant $T>0$ such that $\mathcal {G}\big([t,t+T)\big)$ is strongly connected for any $t\geq0$.

(ii) $\mathcal
{G}_{\sigma(t)}$ is said to be {\it uniformly (jointly) quasi-strongly connected} (UQSC) if there exists a constant $T>0$ such that $\mathcal {G}\big([t,t+T)\big)$ is quasi-strongly connected for any $t\geq0$.

(iii)  Assume that $\mathcal {G}_{\sigma(t)}$ is bidirectional for any $t\geq 0$. $\mathcal
{G}_{\sigma(t)}$ is said to be {\it infinitely jointly connected} (IJC) if $\mathcal {G}\big([
t,\infty)\big)$ is  connected for any $t\geq0$.
\end{defn}

Let a piecewise continuous function
$a_{ij}(t)>0$ be the weight of arc $(j,i), i,j\in
\mathcal {V}$. We study the following agent-dynamics with noise, which was first introduced in \cite{saber04}.
\begin{equation}\label{9}
\dot{x}_i(t)=u_i(t)=\sum\limits_{j \in
\mathcal {N}_i(\sigma(t))}a_{ij}(t)\big(x_j(t)-x_i(t)\big)+w_i(t),\; \; i=1,\dots,N.
\end{equation}
where $w_i(t)$ is a function which describes disturbances.

We have the following assumption on the weight functions $a_{ij}(t)$.

\noindent{\bf A2.} {\it (Weights Rule)} There are two constants $0<a_\ast\leq a^\ast$ such that
$$
a_\ast\leq \int_{t}^{t+\tau_D}a_{ij}(s)ds\leq a^\ast, \ \;t\in \mathds{R}^+.
$$

\begin{remark}
{
Note that assumption A2 is much weaker than the compact assumption,  which is widely used in the literature \cite{jad03,tsi1,tsi2,caoming1,caoming2,shi09,shi11}, requiring that the arc weights are restricted within a compact set, and therefore they always have positive lower and upper bound. Indeed, A2 allows us to deal with general weight functions like $a_{ij}(t)=|\sin t|$ which cannot be covered by the compact assumption. }
\end{remark}
Introduce $\mathcal {F}\doteq \{z:\mathds{R}_{\geq0}\rightarrow \mathds{R}^N: \|z\|_\infty<\infty, $ and $z$ is $C^0$  except for a set with measure 0$\}$.  In order to ensure the existence of the solutions of  (\ref{9}), we impose the following assumption on the regularity of the disturbance function $w(t)=(w_1(t),\dots,w_n(t))^T$.

\noindent{\bf A3.} {\it (Disturbance Regularity)} $w(t)\in \mathcal {F}$.

We assume that assumptions A1-A3 are standing assumptions. Under Assumptions A1 and A2, the set of discontinuity points for the right hand side of equation (\ref{9}) has measure zero. Therefore,  the Caratheodory solutions of (\ref{9}) exist for arbitrary initial conditions, and they are absolutely continuous functions that satisfy (\ref{9}) for almost all $t$ on the maximum interval of existence  \cite{fili, cortes}. In the following, each solution of (\ref{9}) is considered in the sense of Caratheodory without explicit mention.

\subsection{The Robust Consensus Problem}
Consider (\ref{9}) with initial condition $x(t_0)=(x_1(t_0),\dots,x_N(t_0))^T=x^0\in \mathds{R}^N, t_0\geq0$. Let
$$
\hbar(t)\doteq\max_{i\in\mathcal {V}} \{x_i(t)\},\quad \ell(t)\doteq\min_{i\in\mathcal {V}}\{x_i(t)\}
$$
be the maximum and minimum state value  at time $t$, respectively. Denote $\mathcal {H}\big(x(t)\big)=\hbar(t)-\ell(t)$. Inspired by the concepts of input-to-state stability (ISS) and integral input-to-state stability (iISS) \cite{sontag-wang,sontag1}, we introduce the following definition.
\begin{defn}
(i) ($L^\infty$ to $L^\infty$) System (\ref{9}) achieves a global robust consensus (GRC) if there exist a $\mathcal {K}\mathcal {L}$-function
$\beta$ and a $\mathcal {K}$-function $\gamma$ such that for all $w\in \mathcal {F}$ and  initial conditions $x(t_0)=x^0$,
\begin{equation}\label{1}
\mathcal {H}\big(x(t)\big)\leq
\beta\big(\mathcal {H}(x^0),t \big)+\gamma(\|w
\|_{\infty}), \; t\geq0.
\end{equation}

(ii) ($L^2$ to $L^\infty$) System (\ref{9}) achieves a global integral robust consensus (GIRC) if there exist a $\mathcal {K}\mathcal {L}$-function
$\beta$ and a $\mathcal {K}$-function $\gamma$ such that for all $w\in \mathcal {F}$ and initial conditions $x(t_0)=x^0$,
\begin{equation}\label{2}
\mathcal {H}\big(x(t)\big)\leq
\beta\big(\mathcal {H}(x^0),t \big)+\int_0^t \gamma(|w(s)|)ds,  \; t\geq0.
\end{equation}
\end{defn}

We also introduce the following definition  on consensus.

\begin{defn}
 (i) A global consensus (GC) is achieved for system (\ref{9}) if  for any initial condition $x(t_0)=x^0\in \mathds{R}^N$,
 $$\lim_{t\rightarrow \infty}\mathcal {H}\big(x(t)\big)=0$$

(ii) Assume that $\mathcal {F}_0\subseteq \mathcal {F}$. Then a global asymptotic consensus (GAC) with respect to $\mathcal {F}_0$ is achieved for system (\ref{9}) if $\forall x^0\in \mathds{R}^N$, $\forall w\in\mathcal {F}_0$, $\forall \varepsilon>0$, $\forall c>0$, $\exists T>0$ such that $\forall t_0\geq 0$,
 $$\mathcal {H}(x^0)\leq c\quad \Rightarrow \quad\mathcal {H}\big(x(t)\big)\leq \varepsilon, \forall t\geq t_0+T.$$
\end{defn}
%\begin{remark}
%GAC is to say, with bounded initial condition $\mathcal {H}(x^0)$,  $\mathcal {H}\big(x(t)\big)$ not only converges to $0$, but also converge uniformly in $t$ for all $w\in\mathcal {F}_0$ along trajectories of system (\ref{9}).
%\end{remark}

%

\subsection{Main Results}
The target of the paper is to establish proper connectivity conditions of the underlying communication graph which can ensure robust consensus or integral robust consensus.

For networks with general directed communication graphs, we have the following conclusions.

\begin{thm}\label{thm1}
System (\ref{9}) achieves a GRC iff $\mathcal {G}_{\sigma(t)}$ is UQSC.
\end{thm}

\begin{thm}\label{thm2}
System (\ref{9}) achieves a GIRC if $\mathcal {G}_{\sigma(t)}$ is UQSC.
\end{thm}

{
It has been shown in \cite{sontag1} that ISS implies  iISS. Now we see from  Theorems \ref{thm1} and \ref{thm2} that  $$
GRC \Longleftrightarrow  UQSC \Longrightarrow  GIRC.
$$
This leads to the fact that GRC implies GIRC, which is consistent with the ISS and iISS properties.}

Next, when the communication graph is restricted to be bidirectional all the time, we have the following result.
\begin{thm}{\label{thm3}}
Assume that $\mathcal {G}_{\sigma(t)}$ is bidirectional for any $t\geq 0$. System (\ref{9}) achieves a GIRC iff $\mathcal {G}_{\sigma(t)}$ is IJC.
\end{thm}

This robust consensus problem is generally challenging due to the coupled agent dynamics, especially under directed communication and time-varying arc weights. Moreover, a common Lyapunov function is often missing when we consider joint connectivity conditions. To obtain the desired GRC or GIRC inequalities, we have to derive the convergence rate for consensus explicitly against the impact of the disturbance.

We will present the convergence analysis of the main results in Sections 4 and 5, respectively, for directed and bidirectional graphs.

\begin{remark}

In \cite{shi11}, a set tacking problem is studied for multi-agent systems guided by multiple leaders. Set input-to-state stability (SISS) and set integral  input-to-state stability (SiISS) are used to describe the set convergence property. We see that the results obtained in this paper are consistent with the SISS and SiISS analysis in \cite{shi11}.

However, the convergence results  in \cite{shi11} cannot be applied to the model discussed in this paper. Note that leaderless consensus is usually a much harder problem than the leader-follower case, especially under time-varying communication graphs. In leader-follower model, the leader(s) can always be treated as center node and therefore the network has a very special topology. The main difficulty here lies in  that the center node may be different for different time intervals and that its dynamics is also influenced by other nodes.  As will be shown in the following discussions, the symmetry in the structure of $\mathcal{H}(t)$, plays a key role in the convergence analysis.

Additionally, different from  \cite{shi11} and most other existing works, the standing assumption on the weights rule in current paper, A2, is a much weaker condition than the usually applied.
\end{remark}
%%%%%%%%%%%%%%%%%%%%%%%%%%%%%%%%%%%%%%%%%%%%%%%%%%%%%%%%%%%%%%%%%%%%%%%%%%%%%%%%%%%%%%%%%%%%%%%%%%%%%%%%%%%%%%%%%%
\section{Convergence: Directed Graphs}
In this section, we establish convergence analysis for directed graphs.

The necessity statement of Theorem \ref{thm1} follows from a similar argument which was used in \cite{lin07}. Assume that  $\mathcal {G}_{\sigma(t)}$ is not UQSC.  Then for any $T_\ast>0$ there exists $t_\ast\geq 0$ such that
$\mathcal
{G}\big([t_\ast,t_\ast+T_\ast)\big)$ is not quasi-strongly connected.  Consequently,  there exists two distinct nodes $i$ and $j$ such that $\mathcal {V}_1\cap \mathcal {V}_2=\emptyset$, where $\mathcal {V}_1=\{\mbox{nodes\ from\ which\ $i$\ is\ reachable\ in}\  \mathcal
{G}\big([t_\ast,t_\ast+T_\ast)\big)\}$ and  $\mathcal {V}_2=\{\mbox{nodes\ from\ which\ $j$\ is\ reachable\ in}$\ $\mathcal
{G}\big([t_\ast,t_\ast+T_\ast)\big)\}$. Let $w_i(t)\equiv 0$ for $i\in \mathcal {V}_1$ and $w_i(t)\equiv 1$ for $i\in \mathcal {V}_2$ when $t\in[t_\ast,t_\ast+T_\ast]$. Let initial condition $t_0$ be $t_\ast$ with $x_i(t_\ast)=0,\forall i\in \mathcal {V}$. It is not hard to see that $\mathcal {H}(x(t_\ast+T_\ast))=T_\ast$. Hence, GRC cannot be achieved since $T_\ast$ can be arbitrarily large.

%However, to prove the sufficiency part is more complicated. We will first propose some preliminaries on Dini derivatives of $\hbar(t)$ and $\ell(t)$ and UQSC graphs, and then we show that the GRC inequality can be obtained by checking the upper bound of $\mathcal {H}\big(x(t)\big)$ in a time interval which is the union of several ``pasted" subintervals.
\subsection{Key Lemmas}
We first establish the following lemma indicating that the Dini derivative of $\hbar(t)$ is bounded above by $|w(t)|$, and the Dini derivative of $\ell(t)$ is bounded below by $-|w(t)|$.

\begin{lem}\label{lem2}For all $t\geq t_0\geq0$, we have
$$
D^+\hbar(t)|\leq |w(t)|; \quad D^+\ell(t)\geq -|w(t)|
$$

\end{lem}

\noindent{\it Proof.}  We prove $D^+\hbar(t)\leq |w(t)|$. The other part can be proved similarly.

Let $\mathcal{I}(t)$ represent the set containing all the agents
that reach the maximum in the definition of $\hbar(t)$ at
time $t$, i.e.,  $\mathcal{I}(t)=\{i\in\mathcal{V}|\ x_i(t)=\hbar(t)\}$. Then according to Lemma \ref{lem1}, we obtain
\begin{align}
D^+\hbar(t)=\max_{i\in\mathcal{I}(t)} \dot{x}_i(t)= \max_{i\in\mathcal{I}(t)} \Big[\sum\limits_{j \in
N_i(\sigma(t))}a_{ij}(t)\big(x_j(t)-x_i(t)\big)+w_i(t)\Big]\leq \max_{i\in\mathcal{I}(t)}w_i(t)\leq |w(t)|, \nonumber
\end{align}
which completes the proof. \hfill$\square$

We next establish two lemmas on UQSC graphs.

\begin{lem}\label{lem3}
Suppose $\mathcal {G}_{\sigma(t)}$ is UQSC. Then there exists a center $i_0$ from which there is a path $i_0 \rightarrow i$
for all $i\in\mathcal {V}$ in $\mathcal
{G}\big([t,t+\hat{T})\big)$ with $\hat{T}\doteq T+2\tau_D$, and each arc of
$i_0 \rightarrow i$ exists in a time interval with length $\tau_D$ at
least during $[t,t+\hat{T})$.
\end{lem}

 \noindent{\it Proof.} Denote  $t_1$ as the first moment when the
interaction topology switches within $[t,t+\hat{T})$ (to suppose there are
switchings is without loss of generality).

If $t_1\geq t+\tau_D$, then,
 there exists a center $i_0$ from which there is a path $i_0 \rightarrow i$
for all $i\in\mathcal {V}$ in $\mathcal
{G}\big([t,t+T)\big)$ since $\mathcal
{G}\big([t,t+T)\big)$ is quasi-strongly connected, and moreover, each arc of path $i_0 \rightarrow i$ stays there for at least the
dwell time $\tau_D$ during $[t,t+T+\tau_D)$ due to the definition of
$\tau_D$.

 On the other hand, if $t_1<t+\tau_D$, we have
$t_1+T+\tau_D<t+\hat{T}$. Then, for any $i\in\mathcal {V}$, there is
also a  center $i_0$ from which there is a path $i_0 \rightarrow i$
for all $i\in\mathcal {V}$ in $\mathcal
{G}\big([t_1,t_1+T)\big)$, each arc of which exists for at least $\tau_D$ during $[t_1,t_1+T+\tau_D)$. This completes the
proof. \hfill$\square$

Suppose that $\mathcal {G}_{\sigma(t)}$ is UQSC. Define a set-valued function $f:\mathbb{Z}^+\rightarrow 2^{\{1,\dots, N\}}$, where $2^{\{1,\dots, N\}}$ represents the (power)  set containing all the subsets of $\{1,\dots,  N\}$:
$$
f(s)=\{j|j\ \mbox{is\ a\ center\ in\ }\mathcal
{G}\big([(s-1)\hat{T}, s\hat{T})\big)\ \mbox{satisfying the condition of Lemma}\ \ref{lem3}\}, \quad s=1, 2, \dots.
$$

\begin{lem}\label{lem4}Assume that $\mathcal {G}_{\sigma(t)}$ is UQSC and let $d_0$ be the (generalized) diameter of $ \mathcal
{G}\big([0, +\infty)\big)$. Then for any $t=1,2,\dots$, there exists $k_0\in\{1,2,\dots,N\}$ such that
$k_0\in f(s)$ for $s$ as many as at least $d_0$ during $s\in[t, t+(d_0-1)N]$.
\end{lem}

\noindent{\it Proof.} Suppose $k\in f(s)$ for less than $d_0$ times (i.e., less than or equal $d_0-1$) during $[t, t+(d_0-1)N]$ for all $k\in\{1,2,\dots,N\}$. Then, the total number of the elements of all the preimages of $f$ on interval $s\in[t, t+(d_0-1)N]$ is no larger than $(d_0-1)N$. However, on the other hand, there are at least $(d_0-1)N+1$ elements (counting times for the same node) belonging to $f(s)$ during $s\in[t, t+(d_0-1)N]$ since $f(\varsigma)\neq \emptyset$ for all $\varsigma=1,2,\dots$. Then we get the contradiction and the conclusion is proved. \hfill$\square$

%%%%%%%%%%%%%%%%%%%%%%%%%%%%%%%%%%%%%%%%%%%%%%%%%%%%%%%%%%%%%%%%%%%%%%%%%%%%%%%%%%%%%%%%%%%%%%%%%%%%%%%
\subsection{Proof of  Theorem \ref{thm1}: UQSC Implies GRC}
We are now in a position to prove the sufficiency statement in Theorem \ref{thm1}. Assume that the initial time is $t_0=0$ for simplicity. The analysis of $\mathcal {H}\big(x(t)\big)$ will be carried out on time intervals $t\in[sK_0,(s+1)K_0]$ for $s=0,1,2, \dots$, where $K_0=[(d_0-1)N+1]\hat{T}$.

Based on Lemma \ref{lem2}, we see that for all $t\in[sK_0,(s+1)K_0]$,
\begin{equation}\label{12}
\hbar(t)\leq \hbar(sK_0)+\| w \|_\infty  K_0 ; \quad \ell(t)\geq \ell(sK_0)-\| w \|_\infty K_0.
\end{equation}

We divide the rest of the proof into three steps, in which convergence bound will be given over the network node by node on time intervals $[sK_0,(s+ 1 )K_0], s=0,1,\dots$. Assume that
\begin{equation}\label{4}
x_{k_0}(sK_0)\leq \frac{1}{2}\ell(sK_0)+ \frac{1}{2}\hbar(sK_0).
\end{equation}

\noindent{\it Step 1.} According to Proposition \ref{lem4}, we can choose $k_0\in\mathcal{V}$ be the center of $d_0$ joint graphs, $[j_m\hat{T},(j_m+1)\hat{T})\subseteq [sK_0,(s+ 1 )K_0], m=1,2,\dots,d_0$. In this step, we bound $x_{k_0}(t)$ on time interval $[sK_0,(s+1)K_0]$.

 With (\ref{12}), we have
\begin{align}\label{7}
\frac{d}{dt}x_{k_0}(t)
&\leq -\mathcal{Y}_{k_0}(t)\Big(x_{k_0}(t)-\hbar(sK_0)-K_0 \| w \|_\infty \Big)+| w(t) |,\;\ t\in[sK_0,(s+1)K_0]
\end{align}
 where $\mathcal{Y}_{i}(t)=\sum\limits_{j \in
\mathcal {N}_{i}(\sigma(t))}a_{ij}(t)$, which implies
\begin{align}\label{8}
x_{k_0}(t)&\leq \Big[1-e^{-\int_{sK_0}^t\mathcal{Y}_{k_0}(\tau)d\tau} \Big](\hbar(sK_0)+K_0\| w \|_\infty  )+e^{-\int_{sK_0}^t\mathcal{Y}_{k_0}(\tau)d\tau }x_{k_0}(sK_0)+ K_0\|w\|_\infty\nonumber\\
 &\leq \xi_0\ell(sK_0)+(1-\xi_0)\hbar(sK_0)+2K_0\| w \|_\infty, \;\ \ t\in[sK_0,(s+1)K_0]
\end{align}
where $\xi_0= e^{-\lceil\frac{K_0}{\tau_D}\rceil a^\ast (N-1)}/2$ with $\lceil z\rceil$ denoting the smallest integer which is no smaller than $z$. Here the first inequality of (\ref{8}) follows from Gronwall's inequality, and the second one holds based on assumption A2, (\ref{4}) and the simple fact that $\ell(t)\leq \hbar(t)$.

\noindent{\it Step 2.} Since $k_0$ is a center in $\mathcal {G}\big([j_1\hat{T},(j_1+1)\hat{T})\big)$, we can well define a set $\mathcal {V}_1=\{j: \exists  t_1\ s.t.\ (k_0,j)\in\mathcal {G}_{\sigma(t)}$ for $t\in[ t_1, t_1+\tau_D)\subseteq [j_1 \hat{T},(j_1+1)\hat{T}) \}$. In this step, we will establish an upper bound for $x_{k_1}(t), k_1\in\mathcal {V}_1$.

 %There are two cases.
%\begin{itemize}
%\item There exists a time instant $\bar{t}_1\in [ t_1, t_1+\tau_D]$ such that
%\begin{equation}\label{5}
%x_{k_1}(\bar{t}_1)\leq x_{k_0}(\bar{t}_1)\leq \xi_0\ell(sK_0)+(1-\xi_0)\hbar(sK_0)+2K_0\| w \|_\infty.
%\end{equation}
%\item Otherwise,  $x_{k_1}(t)\geq x_{k_0}(t), t\in[ t_1, t_1+\tau_D)$. Thus, one has
We have
\begin{align}
\frac{d}{dt}x_{k_1}(t)&\leq \hat{\mathcal{Y}}_{k_1}(t)\big(\hbar(sK_0)+K_0\| w \|_\infty -x_{k_1}(t)\big)\nonumber\\
&\ \ \ \ +a_{k_1k_0}(t)\Big(\xi_0\ell(sK_0)+(1-\xi_0)\hbar(sK_0)+2K_0\| w \|_\infty-x_{k_1}(t)\Big)+w_{k_1}(t)\nonumber
\end{align}
for $t\in[ t_1, t_1+\tau_D)$, where $\hat{\mathcal{Y}}_{k_1}(t)=\mathcal{Y}_{k_1}(t)- a_{k_1k_0}(t)$. Using  Gronwall's inequality we thus obtain
\begin{align}\label{v1}
x_{k_1}( t_1+\tau_D)&\leq e^{-\int_{ t_1}^{ t_1+\tau_D}\mathcal{Y}_{k_1}(t)dt}x_{k_1}(t_1)+\big(\hbar(sK_0)+K_0\| w \|_\infty\big)\int_{ t_1}^{ t_1+\tau_D} e^{-\int_{ t}^{ t_1+\tau_D} \mathcal{Y}_{k_1}(\tau)d\tau} \hat{\mathcal{Y}}_{k_1}(t) dt\nonumber\\
&\ \ \ +\big(\xi_0\ell(sK_0)+(1-\xi_0)\hbar(sK_0)+2K_0\| w \|_\infty\big)\nonumber\\
&\ \ \ \ \ \  \cdot\int_{ t_1}^{ t_1+\tau_D} e^{-\int_{ t}^{ t_1+\tau_D} \mathcal{Y}_{k_1}(\tau)d\tau} a_{k_1k_0}(t)dt +\tau_D \| w \|_\infty\nonumber\\
&\leq \Big(\xi_0 \int_{ t_1}^{ t_1+\tau_D} e^{-\int_{ t}^{ t_1+\tau_D} \mathcal{Y}_{k_1}(\tau)d\tau} a_{k_1k_0}(t)dt\Big)\ell(sK_0)\nonumber\\
&\ \ +\Big(1-\xi_0 \int_{ t_1}^{ t_1+\tau_D} e^{-\int_{ t}^{ t_1+\tau_D} \mathcal{Y}_{k_1}(\tau)d\tau} a_{k_1k_0}(t)dt\Big)\hbar(sK_0) +(2K_0+\tau_D)\| w \|_\infty,
\end{align}
where the second inequality follows from the facts that $x_{k_1}(t_1)\leq \hbar(sK_0)+K_0\| w \|_\infty$ and
$$
\int_{ t_1}^{ t_1+\tau_D} e^{-\int_{ t}^{ t_1+\tau_D}\mathcal{Y}_{k_1}(\tau)d\tau}\mathcal{Y}_{k_1}(t)dt=1- e^{-\int_{ t_1}^{ t_1+\tau_D}\mathcal{Y}_{k_1}(t)dt}.
$$
Furthermore, noticing that
\begin{align}
\int_{ t_1}^{ t_1+\tau_D} e^{-\int_{ t}^{ t_1+\tau_D} \mathcal{Y}_{k_1}(\tau)d\tau} a_{k_1k_0}(t)dt&=\int_{ t_1}^{ t_1+\tau_D} e^{-\int_{ t}^{ t_1+\tau_D} \hat{\mathcal{Y}}_{k_1}(\tau)d\tau}\cdot e^{-\int_{ t}^{ t_1+\tau_D} a_{k_1k_0}(\tau)d\tau} a_{k_1k_0}(t)dt\nonumber\\
&\geq e^{-(N-2)a^\ast}  \int_{ t_1}^{ t_1+\tau_D} e^{-\int_{ t}^{ t_1+\tau_D} a_{k_1k_0}(\tau)d\tau} a_{k_1k_0}(t)dt\nonumber\\
&= e^{-(N-2)a^\ast}\Big( 1-  e^{-\int_{ t_1}^{ t_1+\tau_D} a_{k_1k_0}(t)dt}\Big)\nonumber\\
&\geq e^{-(N-2)a^\ast}\big( 1-  e^{-a_\ast}\big),
\end{align}
we conclude from (\ref{v1}) that
 \begin{align}\label{6}
x_{k_1}( t_1+\tau_D)
   \leq \zeta\xi_0\ell(sK_0) +( 1-  \zeta\xi_0)\hbar(sK_0) +(2K_0+\tau_D)\| w \|_\infty,
\end{align}
  where $\zeta=e^{-(N-2)a^\ast}( 1-  e^{-a_\ast})$.
%\end{itemize}

Therefore, (\ref{6}) implies that for any $k_1\in\mathcal{V}_1$, there exists an instance $t_\ast\in[j_1 \hat{T},(j_1+1)\hat{T})$ such that
 \begin{align}
x_{k_1}( t_\ast)
   \leq \zeta\xi_0\ell(sK_0) +( 1-  \zeta\xi_0)\hbar(sK_0) +3K_0\| w \|_\infty.
\end{align}
Applying inequality (\ref{7}) on $x_{k_1}$ for $t\in[t_\ast,(s+1)K_0]$, it turns out that
\begin{equation}
x_{k_1}(t)\leq \xi_1\ell(sK_0)+(1-\xi_1)\hbar(sK_0)+4K_0\| w \|_\infty,\;\ t\in[(j_1+1)\hat{T},(s+1) K_0 ]
\end{equation}
for all $k_1\in \mathcal {V}_1$, where $\xi_1= e^{-\lceil\frac{K_0}{\tau_D}\rceil a^\ast (N-1)}e^{-(N-2)a^\ast}( 1-  e^{-a_\ast})  \cdot\xi_0$.
\vspace{2mm}

\noindent{\it Step 3.} Continuing the analysis on time interval $[j_2\hat{T},(j_2+1)\hat{T})$, we can similarly define $\mathcal {V}_2=\{j: \exists  t_2\ s.t.$ there is an arc from $\{k_0\}\cup\mathcal {V}_1$ to $j$  for $t\in[ t_2, t_2+\tau_D)\subseteq [j_2\hat{T},(j_2+1)\hat{T})\}$. Repeating the analysis in Step 2, we have
\begin{equation}
x_{k_2}(t)
\leq \xi_2\ell(sK_0)+(1-\xi_2)\hbar(sK_0)+8K_0\| w \|_\infty,\ \ \ t\in[(j_2+1)\hat{T},(s+ 1 )K_0]
\end{equation}
for all $k_2\in \mathcal {V}_2$, where $\xi_2=e^{-\lceil\frac{K_0}{\tau_D}\rceil a^\ast (N-1)}e^{-(N-2)a^\ast}( 1-  e^{-a_\ast})  \cdot\xi_1$.

 Recall that $d_0$ is the (generalized) diameter of $ \mathcal
{G}\big([0, +\infty)\big)$. We can proceed the analysis on time intervals $[j_m\hat{T},(j_{m+1})\hat{T})$ for $m=3,\dots,d_0$ until we obtain
\begin{equation}
x_{i}\big((s+1)K_0\big)
\leq \xi_{d_0}\ell(sK_0)+(1-\xi_{d_0})\hbar(sK_0)+4d_0K_0\| w \|_\infty, \ \ i=1,\dots,N
\end{equation}
where \begin{align}\label{rate}
\xi_{d_0}=e^{-\lceil\frac{K_0}{\tau_D}\rceil a^\ast (d_0+1)(N-1)}e^{-(N-2)d_0a^\ast}( 1-  e^{-a_\ast})^{d_0}/2.
\end{align}
This leads to
\begin{align}\label{16}
\mathcal {H}\big(x((s+ 1 )K_0)\big)
&\leq \xi_{d_0}\ell(sK_0)+(1-\xi_{d_0})\hbar(sK_0)+4d_0K_0\| w \|_\infty- (\ell(sK_0)-\| w \|_\infty K_0 )\nonumber\\
&=(1-\xi_{d_0}) \mathcal {H}\big(x(sK_0)\big)+(4d_0+1)K_0\| w \|_\infty.
\end{align}

For the opposite case of (\ref{4}) with $x_{k_0}(sK_0)> \frac{1}{2}\ell(sK_0)+\frac{1}{2}\hbar(sK_0)$, we see that (\ref{16}) also holds using a symmetric argument by investigating the lower bound for $\ell((s+1)K_0)$.

Since $s$ is arbitrarily chosen in (\ref{16}), we have
\begin{align}
\mathcal {H}\big(x(nK_0)\big)&\leq(1-\xi_{d_0})^{n} \mathcal {H}(x^0)+\sum_{j=0}^{n-1}(1-\xi_{d_0})^j(4d_0+1)K_0 \| w \|_\infty\nonumber\\
&\leq (1-\xi_{d_0})^{n} \mathcal {H}(x^0)+\frac{(4d_0+1)K_0}{\xi_{d_0}}\cdot\| w \|_\infty\nonumber
\end{align}
for any $n=0,1,2,\dots$. From (\ref{12}), we also know
\begin{equation}
\mathcal {H}\big(x(t)\big)\leq\mathcal {H}\big(x(nK_0)\big)+2K_0\|w\|_\infty,\; t\in[nK_0, (n+1)K_0).
\end{equation}
The desired GRC inequality  is therefore obtained by
\begin{equation}\label{14}
\beta\big(\mathcal {H}(x^0),t \big)=(1-\xi_{d_0})^{\lfloor\frac{t}{K_0}\rfloor}\mathcal {H}(x^0), \quad \gamma(\| w \|_\infty)=(2+\frac{4d_0+1}{\xi_{d_0}})K_0\cdot\| w \|_\infty,
\end{equation}
where $\lfloor\frac{t}{K_0}\rfloor$ denotes the largest integer no greater than $\frac{t}{K_0}$. The proof  is  completed.

%%%%%%%%%%%%%%%%%%%%%%%%%%%%%%%%%%%%%%%%%%%%%%%%%%%%%%%%%%%%%%%%%%%%%%%%%%%%%%%%%%%%%%%%%%

\subsection{Convergence Time}
{  The concept of $\epsilon$-convergence  time, sometimes also called $\epsilon$-computation or $\epsilon$-averaging time, has been introduced for discrete-time consensus algorithms to describe the required steps for the network to reach a certain level of consensus captured by a parameter $\epsilon$, and bounds of this $\epsilon$-convergence  time have been extensively established  in the literature \cite{boyd, caoming2, tsi1, tsi2}.

Now let us introduce the following definition of convergence time for the corresponding continuous-time version (\ref{9}) in the absence of noise. Suppose $w(t)\equiv0$. We define
\begin{align}
T_N(\epsilon)=\sup_{x^0\in \mathds{R}^N,\ \mathcal{H}(x^0)\neq 0}\  \min\Big\{\ t: \frac{\mathcal{H}\big(x(t)\big)}{\mathcal{H}(x^0)}\leq \epsilon \Big\}.
\end{align}

From (\ref{14}), we know that  under UQSC communication graphs,
\begin{align}\label{s9}
\mathcal{H}\big(x(t)\big)\leq (1-\xi_{d_0})^{\lfloor\frac{t}{K_0}\rfloor}\mathcal {H}(x^0)
\leq (1-\xi_{d_0})^{\frac{t}{K_0}-1}\mathcal {H}(x^0)=\frac{1}{1-\xi_{d_0}} e^{- \lambda_0 t} \mathcal {H}(x^0).
\end{align}
where
\begin{equation}\label{jz}
\lambda_0=\frac{1}{K_0}\ln\frac{1}{1-\xi_{d_0}}.
 \end{equation}

 Hence, simple computation leads to
 \begin{align}\label{r7}
 T_N(\epsilon)\leq \frac{\log \big((1-\xi_{d_0})\epsilon\big)^{-1}}{\lambda_0}=O\Big(N{e^{\big(\lceil\frac{K_0}{\tau_D}\rceil a^\ast (d_0+1)+d_0a^\ast\big)N}}\Big)\log \epsilon^{-1}
 \end{align}
where by definition $a_N=O(b_N)$ means that $\lim_{N\rightarrow\infty}\frac{a_N}{b_N}$ is a nonzero  constant.

On the other hand, if $\mathcal {G}_{\sigma(t)}$ is USC, for any $t\geq0$ and any node $k\in\mathcal {V}$, $k$ will be the center of joint graphs on $N-1$ subintervals $[t,t+\hat{T}), \dots, [t+(N-2)\hat{T},t+(N-1)\hat{T})$. Moreover,  the generalized  diameter of $ \mathcal
{G}\big([0, +\infty)\big)$ is exactly $N-1$ for USC graphs.  Therefore, replacing $K_0$ with $K_\ast=(N-1)\hat{T}$ and based on the same analysis as in the proof of Theorem \ref{thm1},  similar GRC inequality under USC graphs can  be given by
$$
\beta\big(\mathcal {H}(x^0),t \big)=(1-\xi^\ast_{d_0})^{\lfloor\frac{t}{K_\ast}\rfloor}\mathcal {H}(x^0), \quad \gamma(\| w \|_\infty)=(2K_\ast+\frac{4N-3}{\xi^\ast_{d_0}})\cdot\| w \|_\infty,
$$
where
$$
\xi_{d_0}^\ast=e^{-\lceil\frac{K_\ast}{\tau_D}\rceil a^\ast N(N-1)}e^{-(N-2)(N-1)a^\ast}( 1-  e^{-a_\ast})^{N-1}/2.
$$

Therefore,  under USC communication graphs, we can similarly obtain
  \begin{align}\label{r8}
 T_N(\epsilon)\leq  O\Big({N( 1-  e^{-a_\ast})^{1-N}e^{(N-1)a^\ast\big(\lceil\frac{K_\ast}{\tau_D}\rceil N +N-2\big)}}\Big)\log \epsilon^{-1}.
 \end{align}

\begin{remark}
Compared to the results for discrete-time consensus dynamics with the same (USC) connectivity condition \cite{tsi1,tsi2}, the convergence time given in (\ref{r8}) is relatively conservative.
Intuitively, the system should achieve faster convergence when the topology is USC compared to when it is UQSC. However, this point is not captured in (\ref{r7}) and (\ref{r8}). The reason for this is that the approach taken in the proof of Theorem \ref{thm1} is targeting particularly UQSC graphs, though it can also be used for USC graphs. We believe that there exist sharper bounds for the convergence time under continuous-time dynamics.
\end{remark}}
\subsection{$L^\infty$-Vanishing Noise}
Consider a set defined by
$$
\mathcal {F}_1\doteq\{z\in \mathcal {F}: \lim_{t\rightarrow \infty}z(t)=0\},
$$
and let  $\mathcal {F}_1^0\subseteq \mathcal {F}_1 $ be a subset with $\lim_{t\rightarrow \infty}\sup_{z\in\mathcal {F}_1^0}|z(t)|=0$. Then the following corollary holds.
\begin{prop}\label{pro0}
(i) System (\ref{9}) achieves a GC for any $w\in \mathcal {F}_1$ if  $\mathcal {G}_{\sigma(t)}$ is UQSC.

(ii) System (\ref{9}) achieves a GAC with respect to $ \mathcal {F}_1^0$ iff $\mathcal {G}_{\sigma(t)}$ is UQSC.
\end{prop}

\noindent{\it Proof.} (i) Suppose $\beta$ and $\gamma$ are defined as (\ref{14}). Let $w_0\in \mathcal {F}_1$ be a fixed function. Then,  $\forall \varepsilon>0$, $\exists T(\varepsilon)>0$ such that $|w_0(t)|<\gamma^{-1}(\varepsilon), \forall  t\geq T(\varepsilon)$. Thus, applying Theorem \ref{thm1} on system (\ref{9}) with $t_0=T(\varepsilon)$, we obtain
\begin{equation}\label{13}
\mathcal {H}\big(x(t)\big)\leq \beta\Big(\mathcal {H}\big(x(T(\varepsilon))\big),t-T(\varepsilon)\Big)+\varepsilon.
\end{equation}
Since $\varepsilon$ can be arbitrarily small, the global consensus follows immediately by taking $t\rightarrow\infty$ in (\ref{13}).

(ii) (Sufficiency.)  Suppose $\beta$ and $\gamma$ are defined as (\ref{14}). Then $\forall \varepsilon>0$, $\exists \tilde{T}(\varepsilon)>0$ such that $|w(t)|\leq\gamma^{-1}(\frac{\varepsilon}{2}), \forall  t\geq \tilde{T}(\varepsilon), \forall w\in  \mathcal {F}_1^0 $. Denoting $\omega^\ast=\sup_{t\in[t_0,\tilde{T}]}\{\sup_{z\in\mathcal {F}_1^0}|z(t)|\}$, there will be two cases.
\begin{itemize}
\item When $t_0< \tilde{T}(\varepsilon)$, one has $\forall t\geq t_0$,
\begin{align}\label{36}
\mathcal {H}\big(x(t)\big)&\leq \beta\Big(\mathcal {H}\big(x(\tilde{T}(\varepsilon))\big),t-\tilde{T}(\varepsilon)\Big)+\frac{\varepsilon}{2}\nonumber\\
&\leq \beta\Big(\beta\big(\mathcal {H}(x^0)+\gamma(\omega^\ast), \tilde{T}(\varepsilon)-t_0\big),t-\tilde{T}(\varepsilon)\Big)+\frac{\varepsilon}{2}\nonumber\\
&\leq \beta\Big(\beta\big(\mathcal {H}(x^0)+\gamma(\omega^\ast), 0\big),t-\tilde{T}(\varepsilon)\Big)+\frac{\varepsilon}{2}.
\end{align}
Furthermore, $\forall c>0$, $\exists {T}_1(c,\tilde{T}(\varepsilon))>0$ such that
$$\beta\Big(\beta\big(c+\gamma(\omega^\ast), 0\big),t-\tilde{T}(\varepsilon)\Big)\leq \frac{\varepsilon}{2}, \forall t\geq{T}_1,
$$
\item When $t_0\geq\tilde{T}(\varepsilon)$, one has $\forall t\geq t_0$,
 \begin{align}
\mathcal {H}\big(x(t)\big)\leq \beta(\mathcal {H}(x^0),t-t_0)+\frac{\varepsilon}{2}.
\end{align}
Then $\forall c>0$, $\exists {T}_2(c)>0$ such that $\beta(\mathcal {H}(x^0),t-t_0)\leq\frac{\varepsilon}{2}, \forall t\geq{T}_2$.
\end{itemize}
Taking $T=\max\{T_1,T_2\}$, we obtain
 $$\mathcal {H}(x^0)\leq c \Rightarrow \mathcal {H}\big(x(t)\big)\leq \varepsilon, \forall t\geq t_0+T, \forall w\in  \mathcal {F}_1^0.$$
Hence the sufficient part is proved.

(Necessity.) Suppose  $\mathcal {G}_{\sigma(t)}$ is not UQSC. Then $\forall\varepsilon>0, \forall T_\ast>0$, $\exists W>0$, such that $|w(t)|\leq \frac{\varepsilon}{2T_\ast}, \forall t\geq W$. Moreover, $\forall T_\ast>0$, $\exists t_\ast>M$ such that $\mathcal
{G}\big([t_\ast,t_\ast+T_\ast)\big)$ is not quasi-strongly connected. Furthermore, we define $\mathcal {V}_1$ and $\mathcal {V}_2$ in the same way as the proof of Theorem \ref{thm1}.
 Let initial condition be $t_0=t_\ast$ with $x_i(t_\ast)=0,\forall i\in \mathcal {V}_1$ and $x_i(t_\ast)=c,\forall i\in \mathcal {V}_2$. Then it is not hard to find that
$\mathcal {H}(x(t_\ast+T_\ast))\geq c-\varepsilon$. Therefore, the global asymptotic consensus cannot be achieved since $T_\ast$ can be arbitrarily large. \hfill$\square$

\subsection{UQSC and GIRC}
\subsubsection{Non-conservativeness}
Theorem 2 only states the sufficiency of UQSC graph for GIRC. Let us see the following simple example which shows that the corresponding necessity claim does not hold.

\noindent {\bf Example 1} Suppose there are only two nodes, $1$ and $2$, in the network. The arc set is $\mathcal {E}_{\sigma(t)}=\{(1,2)\}$ if $t\in [10^k, 10^k+1)$ for $k=0,1,\dots$, and $\mathcal {E}_{\sigma(t)}=\emptyset$ otherwise. Take $a_{12}(t)\equiv a_{21}(t)\equiv 1$. Then we have
\begin{equation}
\frac{d}{dt}\big(x_1(t)-x_2(t)\big)=\begin{cases} -(x_1(t)-x_2(t))+w_1(t)-w_2(t),\;\; t\in [10^k, 10^k+1),\, k=0,1,\dots\\
w_1(t)-w_2(t),\; \;  otherwise\end{cases}
\end{equation}

It is not hard to see that the system in Example 1 achieves a GIRC, though the varying communication graph is not UQSC. Therefore,  UQSC graph is no longer necessary to ensure GIRC with general directed communications.

\subsubsection{Proof of Theorem \ref{thm2}}
The proof follows the same line as the proof of Theorem \ref{thm1}.  We will bound $\mathcal {H}\big(x(t)\big)$ on time intervals $t\in[sK_0,(s+1)K_0]$ for $s=0,1, \dots$.
Denote $\eta_s=\int_{sK_0}^{(s+1)K_0} |w(t)| dt$. Then based on Lemma \ref{lem2}, for any $t\in[sK_0,(s+1)K_0]$, we have
\begin{equation}\label{i3}
x_i(t)\in [\ell(s\hat{T})-\eta_s ,  \hbar(s\hat{T})+\eta_s ], \;i=1,\dots, N.
\end{equation}

Suppose $k_0$ is a  node as defined in the proof of Theorem \ref{thm1}.  Provided, without loss of generality,  that
$x_{k_0}(sK_0)\leq \frac{1}{2}\ell(sK_0)+\frac{1}{2}\hbar(sK_0)$ and as that
\begin{align}
\frac{d}{dt}x_{k_0}(t)
&\leq -\mathcal{Y}_{k_0}(t)\Big(x_{k_0}(t)-\hbar(sK_0)-\eta_s \Big)+| w(t) |,\;\ t\in[sK_0,(s+1)K_0]
\end{align}
 which implies
\begin{align}
x_{k_0}(t)&\leq \Big[1-e^{-\int_{sK_0}^t\mathcal{Y}_{k_0}(\tau)d\tau} \Big](\hbar(sK_0)+\eta_s  )+e^{-\int_{sK_0}^t\mathcal{Y}_{k_0}(\tau)d\tau }x_{k_0}(sK_0)+ \int_{sK_0}^{t} e^{-\int_{sK_0}^z\mathcal{Y}_{k_0}(\tau)d\tau }|w(z)|dz\nonumber\\
 &\leq \xi_0\ell(sK_0)+(1-\xi_0)\hbar(sK_0)+2\eta_s, \;\ \ t\in[sK_0,(s+1)K_0]
\end{align}
where the second inequality follows from the simple fact that $0<e^{-\int_{sK_0}^t\mathcal{Y}_{k_0}(\tau)d\tau }\leq1$.

Therefore, similar to the proof of Theorem \ref{thm1}, the analysis can be carried on node by node for different disjoint intervals and we can eventually arrive at
\begin{align}\label{i16}
\mathcal {H}\big(x((s+ 1 )K_0)\big)
\leq (1-\xi_{d_0}) \mathcal {H}\big(x(sK_0)\big)+(4d_0+1)\eta_s.
\end{align}
Consequently, for any $n=0,1,2,\dots$, it holds that
\begin{align}\label{v2}
\mathcal {H}\big(x(nK_0)\big)&\leq(1-\xi_{d_0})^{n} \mathcal {H}(x^0)+(4d_0+1)\sum_{j=0}^{n-1}(1-\xi_{d_0})^{n-1-j} \eta_j\nonumber\\
&\leq  (1-\xi_{d_0})^{n} \mathcal {H}(x^0)+(4d_0+1)\sum_{j=0}^{n-1} \eta_j
\end{align}
Thus, together with the observation that
\begin{equation}
\mathcal {H}\big(x(t)\big)\leq\mathcal {H}\big(x(nK_0)\big)+\int_{nK_0}^t |w(\tau)|d \tau,\; t\in[nK_0, (n+1)K_0),
\end{equation}
the following GIRC inequality is obtained:
\begin{equation}\label{32}
\mathcal {H}\big(x(t)\big)\leq(1-\xi_{d_0})^{\lfloor\frac{t}{K_0}\rfloor}\mathcal {H}(x^0)+(4d_0+1)\int_0^t |w(\tau)|d \tau.
\end{equation}
This completes the proof. \hfill$\square$
\begin{remark}
A sharper inequality can be obtained based on the first inequality of  (\ref{v2}) that
\begin{equation}\label{v3}
\mathcal {H}\big(x(t)\big)\leq(1-\xi_{d_0})^{\lfloor\frac{t}{K_0}\rfloor}\mathcal {H}(x^0)+(4d_0+1)\int_0^t (1-\xi_{d_0})^{\lfloor\frac{t}{K_0}\rfloor-\lceil \frac{\tau}{K_0}\rceil}|w(\tau)|d \tau,
\end{equation}
which will be useful in the following discussions on distributed event-triggered consensus.
\end{remark}

%It is not hard to see that $\mathcal {G}_{\sigma(t)}$ being  JQSC is necessary to ensure GIRC. However, in general, this $[
%t,\infty)$-joint connectivity cannot guarantee GIRC. An interesting question is what is the proper information exchange assumption of the network to make JQSC assumption sufficient %for GIRC. In the following, we make a characterization for  bidirectional and acyclic communication graphs.
%%%%%%%%%%%%%%%%%%%%%%%%%%%%%%%%%%%%%%%%%%%%%%%%%%%%%%%%%%%%%%%%%%%%%%%%%%%%%%%%%%%%%%%%%%%%%%%%%%%%%%%%%%%%%%%%%%%%%%%%%%%%%%%%%%%%%%%%%%%%
\section{Convergence: Bidirectional Graphs}
This section focuses on the proof of Theorem \ref{thm3}. In what follows of this section, we assume that the communications over the network is bidirectional, i.e., $\mathcal {G}_{\sigma(t)}$ is  bidirectional graph for all $t\geq 0$.

{
\subsection{Time-axis Partition}
We introduce a partition, $0=T_0<T_1<T_2<\dots$, for the time-axis.

Let $T_0=0$. Then  $T_k, k=1,2,\dots,$ can be defined by induction as
\begin{align}
&T_k=\inf\Big\{t\geq T_{k-1}:\mathcal {G}\big([
T_{k-1},t)\big)\mbox{ has a connected spanning subgraph,}\nonumber\\
&\ \ \ \quad \quad \quad \mbox{ each arc of which exists at least } \tau_D \mbox{ time within time interval}\ [
T_{k-1},t)\Big\}. \nonumber
\end{align}
See that when $\mathcal {G}_{\sigma(t)}$ is IJC, $T_k$ is finite for any fixed $k=1,2,\dots$.

We can further define
$$
{J}(t)=\max \{k: t>T_k\}.
$$
Then $J(t)$ characterizes how many times for different proper joint graphs being jointly connected during time interval $[0,t)$.}
\subsection{Proof of Theorem \ref{thm3}}
The necessity part of the conclusion is straightforward, so we just focus on the sufficient part.

Denote $\varpi_0=\int_{T_0}^{T_{d_0}} |w(t)| dt$. Then based on Lemma \ref{lem2}, we have
\begin{equation}
\hbar(t)\leq \hbar(T_0)+ \varpi_0 ; \quad \ell(t)\geq \ell(T_0)-\varpi_0
\end{equation}
for all $T_0\leq t\leq T_{d_0}$.

We divide the following analysis into five steps.

\noindent {\it Step 1.} Take $i_0\in\mathcal{V}$ with $x_{i_0}(T_0)=\ell(T_0)$. Denote $\bar{t}_1$ by
$$
\bar{t}_1\doteq \inf\big\{t\geq T_0:\mbox{at least one different node connects $i_0$ in $\mathcal {G}_{\sigma(t)}$}\big\}.
$$
Since $\mathcal {G}_{\sigma(t)}$ is infinitely jointly connected, we have that $\bar{t}_1+\tau_D\leq T_1$ according to the definition of $T_1$.

Note that, we have
$$
x_{i_0}(\bar{t}_1)\leq \ell(T_0)+\int_{T_0}^{\bar{t}_1}|w(t)|dt,
$$
because no other node is connected to $i_0$ during $[T_0,\bar{t}_1]$. Then similar to (\ref{8}), the following inequality
\begin{align}
\frac{d}{dt}x_{i_0}(t)
&\leq -\mathcal{Y}_{i_0}(t)\Big(x_{i_0}(t)-\hbar(T_0)-\varpi_0 \Big)+| w(t) |,\;\ t\in[T_0,T_{d_0}]
\end{align}
implies
\begin{align}\label{91}
x_{i_0}(t)&\leq \Big[1-e^{-\int_{\bar{t}_1}^t\mathcal{Y}_{i_0}(\tau)d\tau} \Big](\hbar(T_0)+\varpi_0  )+e^{-\int_{\bar{t}_1}^t\mathcal{Y}_{i_0}(\tau)d\tau }x_{i_0}(\bar{t}_1)+ \int_{\bar{t}_1}^{\bar{t}_1+\tau_D}|w(t)|dt\nonumber\\
&\leq e^{-\int_{\bar{t}_1}^t\mathcal{Y}_{i_0}(\tau)d\tau }\ell(\bar{t}_1)+\Big[1-e^{-\int_{\bar{t}_1}^t\mathcal{Y}_{i_0}(\tau)d\tau} \Big]\hbar(T_0)+ \varpi_0+\int_{T_0}^{\bar{t}_1}|w(t)|dt+ \int_{\bar{t}_1}^{T_{d_0}}|w(t)|dt\nonumber\\
 &\leq m_0\ell(sK_0)+(1-m_0)\hbar(sK_0)+2\varpi_0
\end{align}
for all $t\in[T_0,\bar{t}_1+\tau_D]$, where $m_0=e^{-(N-1)a^\ast}$.

\noindent {\it Step 2.} Denote $\bar{\mathcal {V}}_1\doteq\big\{j: j$ is a neighbor of $i_0$ in $\mathcal {G}\big([\bar{t}_1,\bar{t}_1+\tau_D]\big)\big\}$. In this step, we bound $x_{i_1}(t)$ for $i_1\in \bar{\mathcal {V}}_1$ in time interval $[\bar{t}_1,\bar{t}_1+\tau_D]$.

Based on similar analysis with (\ref{6}), we can conclude from (\ref{91}) that
\begin{equation}\label{92}
x_{i_1}(\bar{t}_1+\tau_D)
\leq  \zeta m_0\ell(T_0)+(1- \zeta m_0)\hbar(T_0)+ 2\varpi_0 +\int_{\bar{t}_1}^{\bar{t}_1+\tau_D} |w(\tau)|d \tau,
\end{equation}
where $\zeta=e^{-(N-2)a^\ast}( 1-  e^{-a_\ast})$.

\noindent{\it Step 3.} In this step, let us further discuss the bound of $x_i(t)$ after $\bar{t}_1+\tau_D$ for nodes in $\{i_0\}\bigcup \bar{\mathcal{V}}_1$.

Let us view $T_k$ as a switching instance in the graph signal $\sigma(t)$ for any $k=0,1,\dots$.  Denote $p_1$ as the first switching instance after $\bar{t}_1$. Then $p_1\leq T_1$. There will be two cases:
\begin{itemize}
\item[(i)] If $p_1-\bar{t}_1< 2\tau_D$, we can relax (\ref{91}) and (\ref{92}) on time interval $[T_0,p_1]$. In this case, we have for any $i\in\{i_0\}\cup  \bar{\mathcal {V}}_1 $,
\begin{equation}\label{93}
x_{i}(p_1)\leq \eta_0\ell(T_0)+(1- \eta_0)\hbar(T_0)+ 2\varpi_0+\int_{\bar{t}_1}^{p_1} |w(\tau)|d \tau
\end{equation}
where $\eta_0=e^{-2(N-1)a^\ast}e^{-2(N-2)a^\ast}( 1-  e^{-2a_\ast})=e^{-2(2N-3)a^\ast}( 1-  e^{-2a_\ast})$.

\item[(ii)] If  there is no other node connecting $\{i_0\}\cup\bar{\mathcal {V}}_1$ for $t\in[\bar{t}_1+\tau_D, p)$, applying Lemma \ref{lem2} on the subsystem formed by $\{i_0\}\cup\bar{\mathcal {V}}_1$, we  obtain
    \begin{align}\label{94}
x_{i}(p)
&\leq   \zeta m_0\ell(T_0)+(1- \zeta m_0)\hbar(T_0)+ 2\varpi_0 +\int_{T_0}^{\bar{t}_1+\tau_D} |w(\tau)|d \tau +\int_{\bar{t}_1+\tau_D}^{p} |w(\tau)|d \tau\nonumber\\
&\leq \eta_0\ell(T_0)+(1- \eta_0)\hbar(T_0)+ 2\varpi_0+\int_{T_0}^{p} |w(\tau)|d \tau
\end{align}
for all $i\in\{i_0\}\cup  \bar{\mathcal {V}}_1 $.
\end{itemize}

\begin{figure}[H]
\centerline{\epsfig{figure=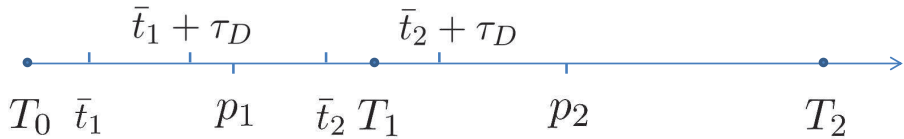, width=0.80\linewidth=0.2}}
\caption{The structure of the proof.}\label{sss}
\end{figure}

\noindent{\it Step 4. } In this step, we continue to analyze the neighbors of $\{i_0\}\cup \bar{\mathcal {V}}_1$. Based on (\ref{93}) and (\ref{94}), if  we  define
$$
\bar{t}_2= \inf\big\{t\geq \bar{t}_1+\tau_D:\ \mbox{at least one other node connects $\{i_0\}\cup \bar{\mathcal {V}}_1$ in $\mathcal {G}_{\sigma(s)}$ for any $s\in[t,t+\tau_D)$}\big\},
$$
and
$$
\bar{\mathcal {V}}_2\doteq\{j: j\mbox{ has a neigbor which belongs to}  \{i_0\}\cup \bar{\mathcal {V}}_1 \mbox{in}\ \mathcal{G}\big([\bar{t}_2,\bar{t}_2+\tau_D)\big)\},
 $$
then we have $\bar{t}_2+\tau_D\leq T_2$ and
\begin{align}
x_{i}(\bar{t}_2)\leq \eta_0\ell(T_0)+(1- \eta_0)\hbar(T_0)+ 3\varpi_0,\ \ \ i\in\{i_0\}\cup  \bar{\mathcal {V}}_1.
\end{align}

Thus, similar to (\ref{92}), a upper bound for any $i_2\in\bar{\mathcal {V}}_2$ can thus be obtained by
\begin{equation}\label{95}
x_{i_2}(\bar{t}_2+\tau_D)
\leq  \zeta m_0\eta_0\ell(T_0)+(1- \zeta m_0\eta_0)\hbar(T_0)+ 5\varpi_0 +\int_{\bar{t}_2}^{\bar{t}_2+\tau_D} |w(\tau)|d \tau,
\end{equation}
and (\ref{95}) also holds for node in $\{i_0\}\cup \bar{\mathcal {V}}_1$. Moreover, if $p_2-\bar{t}_2< 2\tau_D$, where $p_2$ is the first switching instance after $\bar{t}_2+\tau_D$, a relaxed bound can also be obtained as
    \begin{align}
x_{i}(p_2)
&\leq \eta_0^2\ell(T_0)+(1- \eta_0^2)\hbar(T_0)+ 6\varpi_0
\end{align}
for all $i\in\{i_0\}\cup  \bar{\mathcal {V}}_1\cup  \bar{\mathcal {V}}_2 $.

\noindent{\it Step 5.} Proceeding this analysis, $\bar{t}_{3},\dots,\bar{t}_{d_0}$ can be found respectively, and we eventually have
\begin{equation}
x_{i}(\bar{t}_{d_0}+\tau_D)
\leq \eta_0^{d_0}\ell(T_s)+(1-\eta_0^{d_0})\hbar(T_s)+3d_0\varpi_0
\end{equation}
where  $\bar{t}_{d_0}+\tau_D\leq T_{{d_0}}$, which implies
\begin{equation}\label{96}
\mathcal {H}\big(x(T_{d_0})\big)\leq (1-\eta_0^{d_0}) \mathcal {H}\big(x(T_0)\big)+(3d_0+1)\varpi_0.
\end{equation}
Since (\ref{96}) holds independent with the initial condition, we can further conclude that
\begin{equation}
\mathcal {H}\big(x(T_{nd_0})\big)\leq(1-\eta_0^{d_0})^{n} \mathcal {H}(x^0)+(3d_0+1)\sum_{j=0}^{n-1}(1-\eta_0^{d_0})^{n-1-j} \varpi_j
\end{equation}
for all $n=0,1,2,\dots$, where $\varpi_j=\int_{T_{jd_0}}^{T_{(j+1)d_0}}|w(t)|dt$.

Therefore, the desired GIRC inequality can be obtained by
\begin{equation}\label{33}
\mathcal {H}\big(x(t)\big)\leq(1-\eta_0^{d_0})^{\lfloor\frac{{J}(t)}{d_0}\rfloor}\mathcal {H}(x^0)+(3d_0+1)\int_0^t (1-\eta_0^{d_0})^{\lfloor\frac{{J}(t)}{d_0}\rfloor-\lceil\frac{{J}(\tau)}{d_0}\rceil}|w(\tau)|d \tau
\end{equation}

The proof is completed. \hfill$\square$

\subsection{Convergence Time: ``Exponential" Convergence}
{
Suppose $w(t)\equiv0$. Then we see from (\ref{33}) that
\begin{align}\label{68}
\mathcal {H}\big(x(t)\big)\leq (1-\eta_\ast)^{\lfloor\frac{{J}(t)}{d_0}\rfloor}\mathcal {H}(x^0)\leq {\big(1-\eta_\ast\big)^{-1}} e^{-d_0^{-1} {\log (1-\eta_\ast)^{-1}} J(t)} \mathcal {H}(x^0).
\end{align}
where $$\eta_\ast\doteq \eta_0^{d_0}=e^{-(4N-6)d_0a^\ast}( 1-  e^{-2a_\ast})^{d_0}.$$

Now we see from (\ref{68}) that when the system topology  $\mathcal {G}_{\sigma(t)}$ is IJC, system (\ref{9}) with bidirectional communications will reach a consensus exponentially with respect to $J(t)$ in the absence of noise, which is the times of the joint-connection being connected.

Furthermore, an upper bound for the $\epsilon$-convergence time $T_N(\epsilon)$ under IJC communication graph can be established by
\begin{align}
T_N(\epsilon)&\leq \inf \Big\{t: J(t)\leq \frac{d_0}{\log (1-\eta_\ast)^{-1}} \log \big(\epsilon(1-\eta_\ast)\big)^{-1}\Big\}\nonumber\\
&\leq \inf J^{-1}\Big(\big \lceil O \big( e^{(4N-6)d_0a^\ast}\big) \log \epsilon^{-1}\big \rceil\Big)
\end{align}
where $J^{-1}(z)=\{t: J(t)=z\}$.}

\subsection{$L^1$-Vanishing Noise}
Consider the following set:
$$
\mathcal {F}_2\doteq\Big\{z\in\mathcal {F}:\ \int_{0}^\infty |z(t)|dt<\infty\Big\}.
$$
Let $\mathcal {F}_2^0\subseteq \mathcal {F}_2$ be a subset of $\mathcal {F}_2$ with $\int_{0}^\infty \sup_{z\in \mathcal {F}_2^0}|z(t)|dt<\infty$. Then the following corollary holds.
\begin{prop}\label{pro2}
(i) System (\ref{9}) achieves a GAC with respect to $ \mathcal {F}_2^0$ iff $\mathcal {G}_{\sigma(t)}$ is UQSC.

(ii) Assume that  $\mathcal {G}_{\sigma(t)}$ is bidirectional for any $t\geq 0$. Then System (\ref{9}) achieves a GC for all $w\in \mathcal {F}_2$ iff $\mathcal {G}_{\sigma(t)}$ is IJC.
\end{prop}
This proposition follows straightforwardly from the GIRC property given in the previous section. The proof is therefore omitted.

%\begin{lem}Suppose $\{b_j, j=1,2,\dots\}$ is a sequence with $\sum\limits_{j=1}^\infty |b_j|<\infty$ and $0<a<1$. Then we %have $\lim_{n\rightarrow\infty} \sum_{j=1}^n a^{n-j}b_j=0$.
%\end{lem}

\begin{remark} The ideas to obtain Propositions \ref{pro0} and  \ref{pro2} are from ISS and iISS properties which study the conditions to ensure the system state converge to zero \cite{sontag1}. Basically, the results show that to reach a GAC for system (\ref{9}), the system topology has to be UQSC. This is consistent with the main result in \cite{lin07}, in which the problem is studied without disturbances. On the other hand, the results also show that if we require a simple consensus, $[
t,\infty)$-joint connectedness is enough with bidirectional  communications.
\end{remark}

%%%%%%%%%%%%%%%%%%%%%%%%%%%%%%%%%%%%%%%%%%%%%%%%%%%%%%%%%%%%%%%%%%%%%%%%%%%%%%%%%%%%%%%

\section{Application: Distributed Event-triggered Consensus }

Distributed event-triggered coordination for multi-agent systems means that the control input of each agent should be piecewise constant based on neighboring information \cite{dimos,george}.

It has been shown that event-based control needs fewer samples than time-triggered control to achieve the same performance  for stochastic systems \cite{astrom}. Recently event-triggered control has attracted much research interest \cite{tabuada}. Up to distributed event-triggered coordination rules, the system may also benefit from reducing the communication frequency over the network.

We first define the time instances of event-triggered executions for each node. Denote $t^i_0<t^i_1<\dots<t^i_k<\dots$ as the time sequence when agent $i$ is triggered.  Let $t^i_0=0$ be the initial time.  Having got $t_k^i, i\geq 0$, we denote $e_i(t)\doteq x_i(t)-x_i(t^i_k)$  as the error function of node $i$. Let  $L_0$ be a given constant and $\delta (t):\mathds{R}_{\geq0}\rightarrow \mathds{R}_{>0}$ be a given function. Then $t^i_{k+1}$  is determined by the following  triggering condition with forceful waking-up.

(a) Node $i$ keeps checking whether the following equation is satisfied:
\begin{equation}\label{t0}
|e_i(t)|=\delta (t);
\end{equation}

(b) Node $i$ chooses $t^i_{k+1}=t^i_{k}+L_0$ if (\ref{t0}) has never been satisfied over time interval $[t^i_{k},t^i_{k}+L_0]$. Otherwise, $t^i_{k+1}$ equals the first time instance when (\ref{t0}) holds.

\begin{remark}
If the triggering condition is totally determined by equation (\ref{t0}), a node will never be triggered again once its  control input equals zero at some time $t^i_k$.  Consequently, a global consensus cannot be guaranteed. This is why the forceful waking up (timeout) condition is introduced to the triggering condition.
\end{remark}

Next, the communication and updating protocol for the considered distributed event-triggered coordination control is stated as follows.
\begin{itemize}
\item[(i)] {\it (Broadcasting)} Each agent $i$ broadcasts its state $x_i(t^i_k)$ during $[t^i_k, t^i_{k+1})$ until it is triggered another time at $t^i_{k+1}$.
\item[(ii)] {\it (Receiving)}  Agent $j$ can receive $x_i(t^i_k)$ if and only if there exists a time $t_1\in [t^i_k, t^i_{k+1})$ such that $i$ is a neighbor of $j$ at time $t_1$. Moreover, agent $j$ can store this message until another message from $i$ is received.
    \item[(iii)] {\it (Updating)} Each agent $i$ updates its control input at time $x_i(t^i_k)$ once it is triggered, based on the messages it receives from the neighbor set $\hat{\mathcal {N}}_i(k)\doteq \bigcup_{t\in[t^i_{k-1}, t^i_{k})}\mathcal {N}_i(\sigma(t))$.
\end{itemize}

%Furthermore, we define a function $\rho_i(t)$ by
%\begin{equation}
%\rho_i(t)=\begin{cases} 1, \quad if\ \ \hat{t}_{k+1}^i-t^i_{k}\geq\tau_\ast\\
%Sgn (\hat{t}_{k+1}^i-t), \quad otherwise\end{cases}
%\end{equation}
%for $ t\in [t^i_k, t^i_{k+1}), k=1,2,\dots$, where $Sgn (y)=0$ for $y<0$ and $Sgn (y)=1$ for $y\geq0$.

Sticking to rule (i)-(iii), we present the control rule for each node $i$:
\begin{equation}\label{t1}
u_i(t)=\sum\limits_{j \in
\hat{\mathcal {N}}_i(k)}\Big(x_j\big(t^j_{\mathcal {T}^j_i(k)}\big)-x_i\big(t_k^i\big)\Big),\; \; t\in [t^i_k, t^i_{k+1}),
\end{equation}
where  $\mathcal {T}^j_i(k)\doteq  \max_{l}\{l:t_l^j\leq T^\ast_{ij}(k)\}$ with $T^\ast_{ij}(k)\doteq\max_{t}\{t\in [t^i_{k-1}, t^i_{k}):j\in \mathcal {N}_i(\sigma(t))\}$.

Denote $\mathcal {T}_i(t)= \arg\ \max\limits_{l }\{t_l^i| t_l^i\leq t\}, i=1,\dots,N$, and
\begin{equation}\label{v6}
\hat{w}_i(t)=\sum\limits_{j \in
\hat{\mathcal {N}}_i(k)}\big(e_i(t)-e_j(t)\big)+\sum\limits_{j \in
\hat{\mathcal {N}}_i(k)}\Big(x_j\big(t^j_{\mathcal {T}^j_i(k)}\big)-x_j\big(t^j_{\mathcal {T}_j(t)}\big)\Big).
\end{equation}
 Then we can write (\ref{t1}) into the following form:
\begin{equation}\label{tr1}
u_{i}(t)=\sum\limits_{j \in
\hat{\mathcal {N}}_i(k)}\big(x_j(t)-x_i(t)\big)+\hat{w}_i(t).
\end{equation}

Furthermore, denote $\tau^i_{k+1}= t_{k+1}^i-t_{k}^i$ for $k=0, 1, \dots$ and $i=1,\dots, N$ as the difference between two consecutive event time instances, and  let $\tau_0\doteq \min_{i} \inf_{k} \{\tau^i_{k+1}\}$ be their lower bound. Note that, if $\tau_0>0$, the $Zeno$ behavior, which indicates infinite triggering in finite time \cite{jzhang},  is then avoided.

Take
 $$
\xi_\ast=e^{-\lceil\frac{K_0}{\tau_D}\rceil (d_0+1)(N-1)}e^{-(N-2)d_0}( 1-  e^{-1})^{d_0}/2
$$
as the case with $a_\ast=a^\ast=1$ in the definition of $\xi_{d_0}$ in (\ref{rate}). Denote
\begin{equation}
A_0=\frac{1}{1-\xi_\ast}, \quad \theta_0=\frac{1}{K_0}\ln\frac{1}{1-\xi_\ast}.
 \end{equation} The main result on distributed event-triggered coordination is stated as follows.

\begin{thm}Suppose $\delta(t)=A e^{-\theta t}$ with $A>0$ and $0<\theta<\theta_0$. Let $L_0 $ satisfy
\begin{equation}\label{99}
L_0 e^{2\theta L_0}(N-1) \Big[\frac{(N-1)(4d_0+1) A_0 ^2 }{\theta_0-\theta}+1\Big] <\frac{1}{2}.
\end{equation}
Then System (\ref{0}) with control law (\ref{t1}) achieves a GAC with $\tau_0>0$ if $\mathcal {G}_{\sigma(t)}$ is UQSC.

 %(ii) If $0<\theta<\theta_\ast$, and $L_0 $ is chosen to satisfy the following inequality:
%\begin{equation}
%L_0 e^{2\theta L_0}[\frac{(N-1)(4d_0+1)a^\ast c_\ast ^2 }{\theta_\ast-\theta}+1](N-1)a^\ast   <\frac{1}{2}
%\end{equation}
%then System (\ref{0}) with control law (\ref{t1}) achieves a GAC with $\tau_0>0$ if $\mathcal {G}_{\sigma(t)}$ is USC.
\end{thm}

\noindent{\it Proof.}  Define a function
$$
M(t)\doteq \inf \{\tau^i_{k}|t^i_{k}<t, i=1,\dots,N; k=0,1,\dots\}
$$
as the lower bound for the inter-event times before time $t$. $M(t)$ is obviously non-increasing.  Moreover, for any $j\in\hat{\mathcal {N}}_i(k)$, $j$ can be triggered at most $2L_0/M(t)$ times during $t\in[t^i_{k-1}, t^i_{k+1})$ because $|t^i_{k+1}-t^i_{k-1}|\leq 2L_0$. Thus, based on the definition of $\mathcal {T}^j_i(k)$, we see that
%every agent $j\in\hat{\mathcal {N}}_i(k)$ is triggered at most $\frac{L_0}{M(t)}$ times during time interval $[t^i_{k-1}, t^i_{k+1})$ for $ t^i_{k+1}\leq t$. Therefore,
$$
\Big|x_j\big(t^j_{\mathcal {T}^j_i(k)}\big)-x_j\big(t^j_{\mathcal {T}_j(t)}\big)\Big|\leq \frac{2L_0}{M(t)}\delta (t_{k-1}^i)\leq \frac{2L_0}{M(t)}\delta (t)e^{2\theta L_0},\ \ t\in[t^i_{k-1}, t^i_{k+1}).
$$
Then (\ref{v6}) implies
\begin{align}
 |\hat{w}_i(t)|\leq (N-1) \Big[2+\frac{2L_0 e^{2\theta L_0} }{M(t)}\Big]\delta (t),\ \ i=1,\dots,N.
\end{align}

Noting that, for any $t\geq 0$, each arc in $\mathcal {G}_{\sigma(t)}$ will still be kept in the communication graph defined by neighbor sets $\hat{\mathcal {N}}_i(k)$ $i=1,\dots,N; k=1,\dots$. Hence  the dwell time assumption still stands, and therefore, we can conclude from (\ref{v3}) that
\begin{align}\label{v7}
\mathcal {H}\big(x(t)\big)&\leq(1-\xi_\ast)^{\lfloor\frac{t}{K_0}\rfloor}\mathcal {H}(x^0)+(4d_0+1) \int_0^t (1-\xi_\ast)^{\lfloor\frac{t}{K_0}\rfloor-\lceil\frac{\tau}{K_0}\rceil}(N-1)\Big[2+\frac{2L_0e^{2\theta L_0} }{M(\tau)}\Big]\delta(\tau)d \tau\nonumber\\
&\leq A_0 e^{-\theta_0t}\mathcal {H}(x^0)+2(N-1)(4d_0+1)\Big [1+\frac{e^{2\theta L_0} L_0}{M(t)}\Big]A_0^2e^{-\theta_0t}\int_0^t e^{\theta_0\tau}\delta(\tau)d \tau\nonumber\\
&= A_0 e^{-\theta_0t}\mathcal {H}(x^0)+2(N-1)(4d_0+1)\Big [1+\frac{e^{2\theta L_0} L_0}{M(t)}\Big]\frac{A_0^2A}{\theta_0-\theta}\Big(e^{-\theta t}-e^{-\theta_0t}\Big)
\end{align}

{\it Claim.} $\tau_0>0$.

 We prove the claim by contradiction. Assume that $\tau_0=0$. Then we have $\lim_{t\rightarrow\infty}M(t)=0$. Without loss of generality, let us assume that $\tau^i_{k+1}={\delta(t_{k+1}^i)}/{|u_i(t_k^i)|}$ for all $i=1,\dots,N$ and $k=0,1,\dots$. From (\ref{tr1}), we have
 $$
 |u_i(t_k^i)|\leq (N-1)\mathcal {H}(x(t_k^i))+|\hat{w}_i(t_k^i)|\leq (N-1)\mathcal {H}(x(t_k^i))+ (N-1)\Big[2+\frac{2L_0 e^{2\theta L_0} }{M(t^i_k)}\Big]\delta (t_k^i).
 $$  Now we can  further conclude from (\ref{v7}) that
\begin{align}
\tau^i_{k+1}
&\geq \frac{\delta(t_{k}^i)}{(N-1)\mathcal {H}(x(t_k^i))+ \big[2+\frac{2L_0 e^{2\theta L_0} }{M(t^i_k)}\big]\delta (t_k^i)} \cdot e^{-\theta\tau^i_{k+1}}\nonumber\\
&\geq  \frac{A M(t^i_{k})}{\Big[A_0 \mathcal {H}(x^0)+\frac{2(N-1)(4d_0+1)  A_0 ^2 A}{\theta_0-\theta} + 2A \Big]M(t^i_{k})+ 2A\Big[\frac{(N-1)(4d_0+1)  A_0 ^2 }{\theta_0-\theta}+1\Big] L_0 e^{2\theta L_0}   }\nonumber\\
& \ \ \ \times \frac{e^{-\theta\tau^i_{k+1}}}{N-1 }.\nonumber
\end{align}

Now we know that  for any fixed number $0<\mu<1$, there exists $N_1>0$ such that when ${k}>N_1$,
\begin{equation}\label{100}
\tau^i_{k+1}\geq \mu\cdot \frac{M(t^i_{k})}{2L_0 e^{2\theta L_0} (N-1)\Big[\frac{(N-1)(4d_0+1)  A_0 ^2 }{\theta_0-\theta}+1\Big]   }\cdot{e^{-\theta\tau^i_{k+1}}}.
\end{equation}

Since $\tau_0=0$, there has to be $\tau^{i_0}_{k_0+1}\rightarrow 0$ as $k_0$ tends to infinity such that $\tau^{i_0}_{k_0+1}=M(t^{i_0}_{k_0}+\tau^{i_0}_{k_0+1})$. On the other hand, with (\ref{99}), we can choose $\mu$ and $k_0$ sufficiently large to enforce
$$
\mu \cdot\frac{ 1}{2 L_0 e^{2\theta L_0}(N-1) \Big[\frac{(N-1)(4d_0+1)  A_0 ^2 }{\theta_0-\theta}+1\Big]    }\cdot e^{-\theta\tau^{i_0}_{k_0+1}}>1.
$$
As a result, (\ref{100}) will lead to
\begin{equation}
M(t^{i_0}_{k_0}+\tau^{i_0}_{k_0+1})
>  M(t^{i_0}_{k_0}),
\end{equation}
which contradicts the fact that $M(t)$ is non-increasing. The claim is proved.

With $\tau_0>0$, we further obtain
\begin{align}
 |\hat{w}_i(t)|\leq 2(N-1) \Big[1+\frac{L_0 e^{2\theta L_0} }{\tau_0}\Big]\delta (t),
\end{align}
which guarantees GAC for system (\ref{0}) immediately according to Proposition \ref{pro0}. The desired conclusion follows. \hfill $\square$

\section{Conclusions}
This paper focused on the robustness of continuous-time consensus algorithms of single integrator. We provided a precise answer to  how much connectivity is required for the network to agree asymptotically based on noisy communications.

The idea of input-to-state stability and integral input-to-state stability inspired us to our definitions of robust consensus and integral robust consensus. We showed that uniform joint connectivity is critical with respect to robust consensus for general directed graphs; infinite joint connectivity is critical with respect to integral robust consensus.  Upper bounds for the $\epsilon$-convergence time were obtained as a straightforward result of the robustness analysis.

 The results may have many applications since the assumptions we use are quite general. By a generalized integral version of the weight assumption, the dynamics can cover many models in both theoretical and application study.  As an illustration, we studied  distributed event-triggered coordination using the robust consensus inequality.


\begin{thebibliography}{99}

%\bibitem{clark} F. Clarke, Y. Ledyaev, R. Stern, and P.
%Wolenski.
%\newblock{\em Nonsmooth Analysis and Control Theory}.
%\newblock Speringer-Verlag, 1998.

\bibitem{fili} A. F. Filippov. {\em Differential Equations with Discontinuous Righthand Sides}.
Norwell, MA: Kluwer, 1988.

\bibitem{god}
C. Godsil and G. Royle.
\newblock {\em Algebraic Graph Theory.}
\newblock New York: Springer-Verlag, 2001.

\bibitem{dan} J. Danskin. \newblock The theory of max-min, with applications.
{\em SIAM J. Appl. Math.}, vol. 14, 641-664, 1966.


%\bibitem{rou} N. Rouche, P. Habets, and M.
%Laloy.
%\newblock {\em Stability Theory by Liapunov's Direct Method},
%\newblock New York: Springer-Verlag, 1977.

%\bibitem{dan} J. Danskin. \newblock The theory of max-min, with applications,
%{\em SIAM J. Appl. Math.}, vol. 14, 641-664, 1966.



\bibitem{cortes} J. Cort\'{e}s. Discontinuous dynamical systems-a tutorial on solutions, nonsmooth analysis, and stability. {\it IEEE Control Systems Magazine}, vol. 28, no. 3, 36-73, 2008.

    \bibitem{ber}
C. Berge and A. Ghouila-Houri.
\newblock{\em Programming, Games, and
Transportation Networks}.
\newblock John Wiley and Sons, New York, 1965.


\bibitem{vic95}
T. Vicsek, A. Czirok, E. B. Jacob, I. Cohen, and O. Schochet.
\newblock Novel type of phase transitions in a system of self-driven
particles.
\newblock{\em Physical Review Letters}, vol. 75, 1226-1229, 1995.


\bibitem{mar}
S. Martinez, J. Cort\'{e}s, and F. Bullo. \newblock Motion coordination
with distributed information,
\newblock {\em IEEE Control Systems Magazine}, vol. 27, no. 4, 75-88, 2007.

\bibitem{ren} W. Ren and R. Beard. {\em Distributed
Consensus in Multi-vehicle Cooperative Control}. Springer-Verlag,
London, 2008.

\bibitem{ren05} W. Ren and R. Beard. Consensus seeking in multiagent systems under dynamically
changing interaction topologies. {\em IEEE Transactions on Automatic Control}, vol. 50, no. 5, 655-661,
2005.



\bibitem{wanglin1} L. Wang and  L. Guo. Robust consensus and soft control of multi-agent systems with noises. {\em Journal of Systems Science and Complexity}, vol.21, no.3, 406-415, 2008.

      \bibitem{wanglin2} L. Wang and Z. Liu. Robust consensus of multi-agent systems with noise. {\em Science in China, Series F: Information Sciences}, vol. 52, no.  5, 824-834, 2009.

\bibitem{munz} U. M\"{u}nz, A. Papachristodoulou, and F. Allg\"{o}wer. Robust consensus controller design for nonlinear relative degree two multi-agent systems with communication constraints. {\em IEEE Transactions on Automatic Control}, vol. 56, no. 1, 145-151. 2011.

\bibitem{qhui} Q. Hui, W. Haddad, and S.  Bhat. On robust control algorithms for nonlinear network
consensus protocols. {\em American Control Conference, Seattle}, 5062-5067, 2008.

\bibitem{sabertac} R. Olfati-Saber. Flocking for multi-agent dynamic systems:
algorithms and theory. {\em IEEE Trans. Automatic Control}, vol. 51, no. 3,
401-420, 2006.

%\bibitem{rou} N. Rouche, P. Habets, and M.
%Laloy.
%\newblock {\em Stability Theory by Liapunov's Direct Method},
%\newblock New York: Springer-Verlag, 1977.

%\bibitem{caoming} M. Cao, D. A. Spielman and A. S. Morse. A lower bound on convergence of a distributed network consensus algorithm. {\em  IEEE Conference on Decesion and Control},  2356-2361, 2005.

\bibitem{caoming1} M. Cao,  A. S. Morse and B. D. O. Anderson. Reaching a consensus in a dynamically changing
environment: a graphical approach. \newblock {\em SIAM J. Control Optim.}, vol. 47, no. 2, 575-600, 2008.

\bibitem{caoming2} M. Cao,  A. S. Morse and B. D. O. Anderson. Reaching a consensus in a dynamically changing
environment: convergence rates, measurement
delays, and asynchronous events. \newblock {\em SIAM J. Control Optim.}, vol. 47, no. 2, 601-623, 2008.

\bibitem{hong07} Y. Hong, L. Gao, D. Cheng, and J. Hu.
\newblock
Lyapuov-based approach to multi-agent systems with switching jointly
connected interconnection.
\newblock{\em IEEE Trans. Automatic Control}, vol. 52, 943-948, 2007.

\bibitem{cheng} D. Cheng, J. Wang, and X. Hu,
An extension of LaSalle's invariance principle and its application to
multi-agents consensus. {\em IEEE Trans. Automatic Control}, vol. 53,
1765-1770, 2008.

\bibitem{fax} J. Fax and R. Murray. Information flow and cooperative control
of vehicle formations. \newblock {\em IEEE Trans. Automatic
Control}, vol. 49, no. 9, 1465-1476, 2004.

\bibitem{saber04}
R. Olfati-Saber and R. Murray.
\newblock Consensus problems in the networks of agents with switching topology
and time dealys.
\newblock {\em IEEE Trans. Automatic
Control}, vol. 49, no. 9, 1520-1533, 2004.




\bibitem{tantac} H. G. Tanner, A. Jadbabaie, G. J. Pappas. Flocking in fixed and
switching networks. {\em IEEE Trans. Automatic Control}, vol. 52, no. 5,
863-868, 2007.

\bibitem{tan04} H. G. Tanner, G.
Pappas and V. Kumar. Leader-to-formation stability. {\em IEEE Trans. Robot. Autom.}, vol. 20, no. 3, pp. 443-455, 2004.


\bibitem{lwang} F. Xiao and L. Wang. Asynchronous consensus in continuous-time multi-agent systems with switching topology and time-varying delays. {\em IEEE Trans. Automatic Control}, vol. 53, no. 8, 1804-1816, 2008.


\bibitem{jad03}
A. Jadbabaie, J. Lin, and A. S. Morse.
\newblock Coordination of groups of mobile autonomous agents using nearest neighbor rules.
\newblock {\em IEEE Trans. Automatic Control}, vol. 48, no. 6, 988-1001, 2003.

\bibitem{xie} K. You and L. Xie. Network topology and communication data rate for consensusability of discrete-time multi-agent systems. {\em IEEE Transactions on Automatic Control,} vol.56, no.10, 2262-2275, 2011.

\bibitem{shi09} G. Shi and Y. Hong.
Global target aggregation and state agreement of nonlinear
multi-agent systems with switching topologies. {\em Automatica},
vol. 45, 1165-1175, 2009.

\bibitem{shi11} G. Shi, Y. Hong and K. H. Johansson.
Connectivity and set tracking of multi-agent
systems guided by multiple moving leaders. {\em IEEE Trans. Automatic Control}, to appear 2012.


\bibitem{boyd} S. Boyd, A. Ghosh, B. Prabhakar and D. Shah. Randomized Gossip Algorithms. {\it IEEE Trans.
Information Theory}, vol. 52, no. 6, 2508-2530, 2006.

\bibitem{tsi}
J. N. Tsitsiklis, D. Bertsekas, and M. Athans. Distributed asynchronous
deterministic and stochastic gradient optimization algorithms. {\em
IEEE Trans. Automatic Control}, vol. 31, no. 9, 803-812, 1986.

\bibitem{tsi1} A. Olshevsky and J. N. Tsitsiklis. Convergence speed in distributed consensus and averaging. {\em SIAM J. Control Optim.} vol. 48, no. 1, 33-55,  2009.


\bibitem{tsi2} A. Nedi\'{c}, A. Olshevsky, A. Ozdaglar, and J. N. Tsitsiklis. On distributed
averaging algorithms and quantization effects. {\it IEEE Trans.
Automatic Control}, vol. 54, no. 11,  2506-2517, 2009.

\bibitem{sontag1} E. Sontag. Comments on integral variants of ISS. {\em Systems Control Lett.}, vol. 34, no. 1-2, 93-100, 1998

\bibitem{sontag-wang} E. Sontag and Y. Wang. On characterizations of the
input-to-state stability property. {\em Systems Control Lett.},
vol. 24, 351-359, 1995.

\bibitem{lin05}
Z. Lin, B. Francis and M. Maggiore.
\newblock Necessary and sufficient graphical conditions for formation
control of unicycles.
\newblock{\em  IEEE Trans. Automatic Control}, vol. 50, no. 1,
121-127, 2005.

\bibitem{lin07} Z. Lin, B. Francis, and M. Maggiore.
\newblock State agreement for continuous-time coupled nonlinear systems.
\newblock {\em SIAM J. Control Optim.}, vol. 46, no. 1, 288-307, 2007.

%\bibitem{linwang} Lin Wang and  Lei Guo, Robust Consensus of Multi-Agent Systems under
%Directed Information Exchanges, {\em Chinese Control Conference}, 557-561, 2007


\bibitem{} Y. Zhang and Y.-P Tian. Consentability and protocol design of multi-agent systems with stochastic switching topology. {\em Automatica}, vol. 45, 1195-1201, 2009.



\bibitem{} W. Yu, G. Chen, M. Cao. Some necessary and sufficient conditions for second-order consensus in
multi-agent dynamical systems. {\em Automatica}, vol.46, 1089-1095, 2010.

\bibitem{carli} R. Carli, A. Chiuso,
L. Schenato and S. Zampieri. Optimal synchronization for networks of  noisy double
integrators. {\em IEEE Trans. Automatic Control},  vol. 56, no. 5,
1146-1152, 2011.

\bibitem{Tian}Y.-P Tian and C.-L Liu. Robust consensus of multi-agent systems with diverse input delays and asymmetric interconnection perturbations.{\em  Automatica},  vol. 45(5), 1347-1353, 2009.

\bibitem{leonard} G. F. Young, L. Scardovi and N. E. Leonard. Robustness of noisy consensus dynamics with directed communication. {\em Proc. of the American Control Conf.}, 6312-6317,   2010.

\bibitem{pesco} L. Pescosolido, S. Barbarossa and G. Scutari. Average consensus algorithms robust against channel noise. {\em SPAWC}, 261-265, 2008.

\bibitem{mor}
L. Moreau. Stability of multiagent systems with time-dependent
communication links. {\em IEEE Trans. Automatic Control},  vol. 50, no. 2,
169-182, 2005.

\bibitem{astrom} K. J.  {\AA}str\"{o}m and B. Bernhardsson. Comparison of Riemann and
Lebesgue sampling for first order stochastic systems. {\em  IEEE
Conference on Decision and Control}, 2011-2016, 2002.

\bibitem{lemmon} X. Wang and M. Lemmon. Event-triggering in distributed networked systems with data
dropouts and delays. {\em Hybrid Systems: Computation and Control}, 366-380, 2009.

\bibitem{tabuada} P. Tabuada. Event-triggered real-time scheduling of stabilizing control tasks. {\em IEEE
Transactions on Automatic Control}, vol. 52, no. 9, 1680-1685, 2007.

\bibitem{jzhang} J. Zhang, K. H.  Johansson, J. Lygeros, and S. Sastry. Zeno hybrid systems. {\em International
Journal of Robust and Nonlinear Control}, vol. 11, no. 5, 435-451, 2001.

\bibitem{dimos} D. Dimarogonas and K. H. Johansson. Event-triggered control for multi-agent systems. {\em IEEE Conference on Decision and Control}, Shanghai, China, 7131-7136, 2009.

\bibitem{george} G. Seyboth, D. Dimarogonas and K. H. Johansson. Control of multi-agent systems via
event-based communication, {\em IFAC World Congress},   Milan, 2011.

\end{thebibliography}
\end{document}